\documentclass[camera,letterpaper,nomarginnotes,nonarrowgutter]{jpaper}

\PassOptionsToPackage{table}{xcolor}
\usepackage[sort,compress]{cite}
\usepackage{amsmath,amsfonts}
\usepackage{tikz}
\usepackage{graphicx}
\usepackage{booktabs}
\usepackage{textcomp}
\usepackage{xcolor}
\usepackage{booktabs}
\usepackage{pifont}
\usepackage{siunitx}
\usepackage{tabularray}
\usepackage{adjustbox}
\usepackage{amsmath}
\usepackage{algorithm2e}
\usepackage{multirow}
\usepackage{fancyhdr}
\usepackage{float}
\usepackage{setspace}
\usepackage{listings}
\usepackage{balance}
\usepackage{wrapfig}
\usepackage{marginnote}
\usepackage{boldline} 
\usepackage{colortbl} 
\usepackage{floatflt}
\usepackage{listing}
\usepackage{eso-pic}
\usepackage{datetime}
\usepackage{pifont} 
\usepackage[framemethod=tikz]{mdframed}
\usepackage{xspace}
\usepackage{color}
\usepackage{listings}
\usepackage{stmaryrd}   

\usepackage[bottom]{footmisc}
\usepackage{dblfloatfix}
\usepackage{array}
\usepackage{latexsym}
\usepackage{ifthen}
\usepackage{color}
\usepackage{booktabs}
\usepackage{listings}
\usepackage[most]{tcolorbox}
\usepackage{listings}
\usepackage{chngcntr}
\usepackage[T1]{fontenc} 
\usepackage{zi4} 
\usepackage{float} 
\usepackage[disable,colorinlistoftodos,prependcaption,textsize=scriptsize]{todonotes}
\usepackage{tikz}
\usepackage{hyperref}

\definecolor{urlblue}{HTML}{319dd6}

\newcounter{CurrentDraftVersion}

\newif\ifarxiv
\arxivtrue

\newif\ifsubmission
\submissiontrue

\ifsubmission
    \fancypagestyle{firstpage}
    {
        \fancyhead{}
        
        \pagenumbering{arabic}
        \fancyfoot[C]{\large\thepage}
    }
\else
    \fancypagestyle{firstpage}
    {
        \fancyhead[C]{\textcolor{blue}{\emph{Version \arabic{CurrentDraftVersion}~---~\today, \xxivtime \ UTC}}}
        \fancyfoot[C]{\thepage}
    }
\fi

\hypersetup{
    bookmarks=true,
    breaklinks=true,
    colorlinks=true,
    linkcolor=blue,
    citecolor=blue,
    urlcolor=urlblue,
    pdfpagelabels=false,
}

\definecolor{structuralbg}{RGB}{221, 235, 247}   
\definecolor{mappingbg}{RGB}{252, 228, 214}      
\definecolor{isolationbg}{RGB}{226, 239, 218}    
\definecolor{integritybg}{RGB}{255, 242, 204}    

\definecolor{darkspringgreen}{rgb}{0.09, 0.45, 0.27}
\definecolor{denim}{rgb}{0.08, 0.38, 0.74}
\definecolor{darkolivegreen}{rgb}{0.33, 0.42, 0.18}
\definecolor{tangerine}{rgb}{0.95, 0.52, 0.0}
\definecolor{mahogany}{rgb}{0.75, 0.25, 0.0}
\definecolor{coolblack}{rgb}{0.0, 0.18, 0.39}
\definecolor{yellow}{rgb}{1.1, 0.85, 0.1}
\definecolor{darkblue}{rgb}{0.0, 0.0, 0.7} 
\definecolor{lightmahogany}{rgb}{0.87, 0.45, 0.33}
\definecolor{cerulean}{rgb}{0.0, 0.48, 0.65}
\definecolor{violet}{rgb}{0.56, 0.0, 1.0}
\definecolor{lightorange}{rgb}{1.0, 0.9, 0.8}

\definecolor{darkspringgreen}{rgb}{0.09, 0.45, 0.27}
\definecolor{denim}{rgb}{0.08, 0.38, 0.74}
\definecolor{darkolivegreen}{rgb}{0.33, 0.42, 0.18}
\definecolor{tangerine}{rgb}{0.95, 0.52, 0.0}
\definecolor{mahogany}{rgb}{0.75, 0.25, 0.0}
\definecolor{red}{rgb}{1.0, 0.0, 0.0}

\definecolor{uglyyellow}{rgb}{0.99, 0.93, 0.0}
\definecolor{nbs}{rgb}{0.35, 0.31, 0.81}

\newcommand{\nbc}[1]{\todo[size=\scriptsize, linecolor=black, bordercolor=black, backgroundcolor=white]{\textcolor{nbs}{\textbf{@nb:} #1}}}
\definecolor{darkred}{rgb}{0.75, 0.1, 0.1}

\newcommand{\dbbc}[1]{\todo[size=\scriptsize, linecolor=black, bordercolor=black, backgroundcolor=white]{\textcolor{darkred}{\textbf{@dbb:} #1}}}

\newcommand\hmc[1]{{\color{darkolivegreen}{\fbox{\textsc{\textbf{Harun:}}} \textit{#1}}}}

\newif\ifshowcomments
\showcommentstrue   

\ifshowcomments
  \newcommand{\agycomment}[1]{%
    \todo[size=\scriptsize,
      linecolor=black,
      bordercolor=black,
      backgroundcolor=white]{\scriptsize\textcolor{red}{\textbf{@gy:} #1}}}
\else
  \newcommand{\agycomment}[1]{}
\fi

\showcommentsfalse 
\ifshowcomments
  \newcommand{\fuecomment}[1]{%
    \todo[size=\scriptsize,
      linecolor=black,
      bordercolor=black,
      backgroundcolor=white]{\scriptsize\textcolor{blue}{\textbf{@fe:} #1}}}
\else
  \newcommand{\fuecomment}[1]{}
\fi

\ifshowcomments
  
\else
  \renewcommand{\agycomment}[1]{}
\fi

\newcommand{\squishlist}{
 \begin{list}{$\circ$}
  { \setlength{\itemsep}{0pt}
     \setlength{\parsep}{0pt}
     \setlength{\topsep}{3pt}
     \setlength{\partopsep}{0pt}
     \setlength{\leftmargin}{1em}
     \setlength{\labelwidth}{1em}
     \setlength{\labelsep}{0.5em} } }

\newcommand{\squishend}{
  \end{list}  }
  
\newcommand*\circled[1]{\tikz[baseline=(char.base)]{\node[shape=circle,fill,inner sep=0.5pt] (char) {\textcolor{white}{#1}};}}

\newcounter{take}
\newcounter{fea}
\definecolor{obspink}{RGB}{255,220,240}
\newcounter{obs}

\newmdenv[
  skipabove=4pt,
  skipbelow=2pt,
  leftmargin=0pt,
  rightmargin=0pt,
  innerleftmargin=2pt,
  innerrightmargin=2pt,
  innertopmargin=2pt,
  innerbottommargin=3pt,
  linewidth=0.5pt,
  linecolor=black,
  roundcorner=2pt,
]{observationbox}

\newcommand{\observation}[1]{%
    \refstepcounter{obs}%
    \begin{observationbox}
        {\setlength{\fboxsep}{2pt}%
         \colorbox{obspink}{\textbf{Obs~\#\theobs.}}}%
        \hspace{0.7em}#1
    \end{observationbox}
}

\newtcblisting{mycode}[1][]{
  colback=gray!10,
  sharp corners,
  colframe=gray!10,
  boxrule=0pt,
  left=0.2em, right=0.2em, top=0.2em, bottom=0.2em,
  before skip=0.2em, after skip=0.2em,
  fontupper=\scriptsize,     
  listing only,
  listing options={
    basicstyle=\ttfamily,
    xleftmargin=0.5em, xrightmargin=0.5em,
    aboveskip=0.3pt, belowskip=0.3pt,
    #1
  }
}

\newtcolorbox{mymathbox}[1][]{
  colback=gray!10,
  sharp corners,
  colframe=gray!10,
  boxrule=0pt,
  left=0.2em, right=0.2em,
  top=0.1em, bottom=0.1em,   
  before skip=0.2em, after skip=0.2em,
  fontupper=\scriptsize,     
  #1
}

\definecolor{ufogreen}{rgb}{0.1, 0.6, 0.4}
\definecolor{ufogreen}{rgb}{0.1, 0.6, 0.4}

\lstdefinestyle{custompseudocode}
{
    language=Java,
    basicstyle=\footnotesize\ttfamily,
    commentstyle=\ttfamily\itshape\color{gray},
    stringstyle=\ttfamily,
    showstringspaces=false,
    breaklines=true,
    belowcaptionskip=0.3\baselineskip, 
    frameround=ffff,
    frame=single,
    rulecolor=\color{black},
    tabsize=1,
    keywordstyle=\color{red}\bfseries,
    columns=fullflexible,
    morekeywords={public, class}
    morekeywords={function}, 
    keywordstyle=[2]\bfseries, 
    escapeinside={*@}{@*}, 
}
\lstdefinestyle{customc}{
  belowcaptionskip=1\baselineskip,
  breaklines=true,
  frame=single,
  xleftmargin=\parindent,
  language=C,
  showstringspaces=false,
  basicstyle=\footnotesize\ttfamily,
  keywordstyle=\bfseries\color{green!40!black},
  commentstyle=\itshape\color{tangerine},
  identifierstyle=\color{black},
  stringstyle=\color{orange},
  numbers=left,
  stepnumber=1,
  numbersep=-5pt,
  escapeinside={*@}{@*}, 
}

\definecolor{cadmiumgreen}{rgb}{0.0, 0.42, 0.24}
\definecolor{chocolate}{rgb}{0.92, 0.41, 0.12}
\definecolor{burgundy}{rgb}{0.5, 0.0, 0.13}
\definecolor{darkmagenta}{rgb}{0.55, 0.0, 0.55}
\definecolor{darkblue}{rgb}{0.0, 0.5, 1.0}
\definecolor{lavender}{RGB}{150, 120, 190}

\renewcommand{\nbc}[1]{}
\renewcommand{\dbbc}[1]{}

\renewcommand\hmc[1]{}

\usepackage{mathtools}

\makeatletter

\providecommand{\HECiteCmd}{\cite}

\newcommand{\HE@cite}[1]{\HECiteCmd{#1}}

\newcommand{\citecat}[2][\HECiteCmd]{%
  \begingroup
    \let\HECiteCmd#1\relax
    \@nameuse{HE@cat@#2}%
  \endgroup
}

\newcommand{\HEdefcat}[2]{%
  \expandafter\newcommand\csname HE@cat@#1\endcsname{\HE@cite{#2}}%
}

\HEdefcat{foundations}{%
  RAD78PrivacyHomo,Regev09LWE,LPR10RLWE,Gentry09FHE,%
  Gentry09Thesis,BV11LWEFHE,BGV12LeveledFHE,%
  SV14FHESIMD,GSW13,HS20HElibDesign,DGHV10,Brakerski12ScaleInv,%
  GH11Implementing,AAUC18Survey,Peikert16Decade, rivest1978method, elgamal1985public, paillier1999public, ogburn2013homomorphic, tourky2016homomorphic, yi2014homomorphic
}

\HEdefcat{PUM}{seshadri.bookchapter17,seshadri2013rowclone,chang.hpca16,
kevinchang-thesis,seshadri.thesis16,Seshadri:2015:ANDOR,seshadri.arxiv16, seshadri.micro17,hajinazarsimdram, seshadri2020indram,chi2016prime, shafiee2016isaac,li.micro17, seshadri.bookchapter17.arxiv, deng.dac2018, xin2020elp2im,song2018graphr, song2017pipelayer,gao2020computedram, eckert2018neural,aga.hpca17,fujiki2019duality,besta2021sisa_micro,li.dac16,ferreira2022pluto,imani2019floatpim,he2020sparse,flashcosmos,truong2022adapting,truong2021racer,olgun2021quactrng,kim.hpca19,kim.hpca18,bostanci2022dr,olgun2022pidram,ali2019memory,angizi2019graphide,li2018scope,subramaniyan2017parallel,zha2020hyper,fujiki2018memory,orosa2021codic,sharad2013ultra,gao2021parabit,choi2020flash,han2019novel,merrikh2017high,wang2018three,lue2019optimal,kim2021behemoth,wang2022memcore,han2021flash,kang2021s,lee2020neuromorphic,lee20223d,si2019dual,simon2020blade,nag2019gencache,wang2019bit,al2020towards,kang.icassp14,kim2021colonnade,jiang2020c3sram,jeloka201628,wang2023infinity,kang2015energy,imani2020dual,deng2019lacc,sutradhar2021look,sutradhar2020ppim,peng2023chopper,shahroodi2023swordfish,mimdram,missingnot,yuksel2024simultaneous,yavits2023drama,jahshan2024majork,khalifa2023clapim,garzon2022aida,hanhan2022edam,morad2016resistive,wu2022dramcam_generalpurpose,sadredini2020flexamata,sadredini2019eap,angstadt2018aspen,wang2016sequential,wang2015association}

\HEdefcat{PNM}{fernandez2020natsa,cali2020genasm,kim.bmc18,ahn.pei.isca15,ahn.tesseract.isca15,boroumand.asplos18, boroumand2021google,boroumand2019conda,boroumand2016pim,boroumand.arxiv17,singh2019napel,asghari-moghaddam.micro16,JAFAR,chi2016prime,farmahini2015nda,gao.pact15,DBLP:conf/hpca/GaoK16,gu.isca16,guo2014wondp,hashemi.isca16,cont-runahead,hassan.memsys15,hsieh.isca16,kim.isca16,kim.sc17,DBLP:conf/IEEEpact/LeeSK15,liu-spaa17,morad.taco15,nai2017graphpim,pattnaik.pact16,pugsley2014ndc,zhang.hpdc14,zhu2013accelerating,DBLP:conf/isca/AkinFH15,gao2017tetris,drumond2017mondrian,dai2018graphh,zhang2018graphp,huang2020heterogeneous,zhuo2019graphq,herruzo2021enabling,boroumand-icde-2022, syncron, besta2021sisa_micro, asgarifafnir, upmem2018, shin2018mcdram, cho2020mcdram, denzler2021casper, gomez2022machine, giannoula2022towards, fernandez2022exploiting, oliveira2022heterogeneous, balasubramonian2014near, jacob2016compiling, nair2015active, lloyd2018dse, gokhale2015rearr, lloyd2015memory, rodrigues2016scattergather, lloyd2017keyvalue, landgraf2021combining, nair2015evolution, kwon202125, lee2021hardware, kim2021aquabolt, ke2021near, lee2022improving,siddique2024architectural,jaiyeoba2023acts,lenjani2022pulley,lenjani2021supporting,zhou2021ultra,sadredini2021sunder,lenjani2020fulcrum,mosanu2022pimulator}

\HEdefcat{PIM}{fernandez2020natsa,cali2020genasm,kim.bmc18,ahn.pei.isca15,ahn.tesseract.isca15,boroumand.asplos18,boroumand2021google,boroumand2019conda,boroumand2016pim,boroumand.arxiv17,singh2019napel,asghari-moghaddam.micro16,JAFAR,chi2016prime,farmahini2015nda,gao.pact15,DBLP:conf/hpca/GaoK16,gu.isca16,guo2014wondp,hashemi.isca16,cont-runahead,hassan.memsys15,hsieh.isca16,kim.isca16,kim.sc17,DBLP:conf/IEEEpact/LeeSK15,liu-spaa17,morad.taco15,nai2017graphpim,pattnaik.pact16,pugsley2014ndc,zhang.hpdc14,zhu2013accelerating,DBLP:conf/isca/AkinFH15,gao2017tetris,drumond2017mondrian,dai2018graphh,zhang2018graphp,huang2020heterogeneous,zhuo2019graphq,herruzo2021enabling,boroumand-icde-2022,syncron,besta2021sisa_micro,asgarifafnir,upmem2018,shin2018mcdram,cho2020mcdram,denzler2021casper,gomez2022machine,giannoula2022towards,fernandez2022exploiting,oliveira2022heterogeneous,balasubramonian2014near,jacob2016compiling,nair2015active,lloyd2018dse,gokhale2015rearr,lloyd2015memory,rodrigues2016scattergather,lloyd2017keyvalue,landgraf2021combining,nair2015evolution,kwon202125,lee2021hardware,kim2021aquabolt,ke2021near,lee2022improving,siddique2024architectural,jaiyeoba2023acts,lenjani2022pulley,lenjani2021supporting,zhou2021ultra,sadredini2021sunder,lenjani2020fulcrum,mosanu2022pimulator,seshadri.bookchapter17,seshadri2013rowclone,chang.hpca16,kevinchang-thesis,seshadri.thesis16,Seshadri:2015:ANDOR,seshadri.arxiv16,seshadri.micro17,hajinazarsimdram,seshadri2020indram,shafiee2016isaac,li.micro17,seshadri.bookchapter17.arxiv,deng.dac2018,xin2020elp2im,song2018graphr,song2017pipelayer,gao2020computedram,eckert2018neural,aga.hpca17,fujiki2019duality,li.dac16,ferreira2022pluto,imani2019floatpim,he2020sparse,flashcosmos,truong2022adapting,truong2021racer,olgun2021quactrng,kim.hpca19,kim.hpca18,bostanci2022dr,olgun2022pidram,ali2019memory,angizi2019graphide,li2018scope,subramaniyan2017parallel,zha2020hyper,fujiki2018memory,orosa2021codic,sharad2013ultra,gao2021parabit,choi2020flash,han2019novel,merrikh2017high,wang2018three,lue2019optimal,kim2021behemoth,wang2022memcore,han2021flash,kang2021s,lee2020neuromorphic,lee20223d,si2019dual,simon2020blade,nag2019gencache,wang2019bit,al2020towards,kang.icassp14,kim2021colonnade,jiang2020c3sram,jeloka201628,wang2023infinity,kang2015energy,imani2020dual,deng2019lacc,sutradhar2021look,sutradhar2020ppim,peng2023chopper,shahroodi2023swordfish,mimdram,missingnot,yuksel2024simultaneous,yavits2023drama,jahshan2024majork,khalifa2023clapim,garzon2022aida,hanhan2022edam,morad2016resistive,wu2022dramcam_generalpurpose,sadredini2020flexamata,sadredini2019eap,angstadt2018aspen,wang2016sequential,wang2015association}

\HEdefcat{membottleneck}{mutlu.imw13, mutlu.superfri15, dean2013tail, kanev.isca15, ferdman.asplos12, wang.hpca14, boroumand.asplos18, boroumand2021google, mutlu2019enabling, mutlu2019, ghose2019processing, alser2020accelerating, cali2020genasm, koppula2019eden, kanellopoulos2019smash, oliveira2021pimbench, narancic2014evaluating, ayers2018memory, jia2016understanding, Manegold_2000, gholami2020ai, stengel2015quantifying, chishti2019memory, burger1995declining, gupta2020architectural, sriraman2019softsku, zhao2022understanding, ruan2019insider, gan2018architectural, sriraman2020accelerometer, yuan2023rambda, delimitrou2018amdahl, hsia2020cross, lottarini2018vbench, wang2022characterizing, richins2020missing, mckee.cf04}

\HEdefcat{schemes}{%
  BV11RLWE,FV12BFV,CKKS17,CHKKS18Bootstrap,CHKKS18FullRNS,%
  CCS19ImprovedBootstrap,DM15FHEW,CGGI16FasterBootstrap,CGGI20TFHE,%
  HS15HElibBoot,MP21BootstrappingFHEW%
}

\HEdefcat{libraries}{%
  OpenFHE22,SEAL,ChenLainePlayer17SEAL,HElibRepo,TFHELib,Concrete20,%
  TFHErs,Lattigo,TenSEAL21,IbarrondoViand21Pyfhel,PhantomFHE23,%
  PALISADE,HEIR%
}

\HEdefcat{compilers}{%
  CHET19PLDI,EVA20PLDI,nGraphHE19,nGraphHE2_19,Cingulata15,%
  Porcupine21PLDI,HECO23,Fhelipe24PLDI,ViandJattkeHithnawi21SoKCompilers,%
  Marble18%
}

\HEdefcat{ml}{%
  gilad2016cryptonets,LoLa19,Gazelle18,Delphi20,reagen2021cheetah,%
  LeeMPCNN22,ConcreteML%
}

\HEdefcat{databases}{%
  HE3DB23,ArcEDB24,HEDA22,kabra2025ciphermatch,SealPIR18,OnionPIR21,Spiral22,%
  SimplePIR23,ChenLaineRindal17PSI%
}

\HEdefcat{workloads}{%
  KimEtAl21Genotype,LauterNaehrigVaikuntanathan11,GraepelLauterNaehrig12,%
  CryptoGCN22,SavPoseidon21%
}

\HEdefcat{algorithms}{%
  HS14AlgorithmsHElib,HS18FasterLin,Bossuat21EUROCRYPT,%
  IliashenkoZucca21PoPETs,HalevPolyakovShoup19,KimPolyakovZucca21,%
  ChillottiCGGI17Packed%
}

\HEdefcat{gpu}{%
  DaiSunar15cuHE,NuFHE,Jung21CKKS100x,TensorFHE23HPCA,PhantomFHE23,%
  Shen23CARM,Badawi18GPUFV%
}

\HEdefcat{fpga}{%
  RoyTuran19HPCA,HEAX20,TuranRoy20HEAWS,hexl,FAB23HPCA,%
  Poseidon23HPCA,FxHENN23HPCA,FAST25FPGA,MAD23%
}

\HEdefcat{asic}{%
  F121MICRO,CraterLake22ISCA,BTS22ISCA,ARK22MICRO,SHARP23ISCA,%
  BASALISC22,SoKFHEAccel24%
}

\HEdefcat{pim}{%
  CryptoPIM20,gupta2022memfhe,CiMHE19,gupta2023evaluating,%
  zhou2025fhemem,chen2024hmc,park2025fhendi,kim2025anaheim,mwaisela2024evaluating%
}

\HEdefcat{storage}{%
  CIPHERMATCH25,FlashCosmos22,ParaBit21,AresFlash24,TawoseHE23%
}

\HEdefcat{pet}{%
  PipeZK21ISCA,SecureML17,ABY3_18%
}

\HEdefcat{surveys}{%
  HEStandard18,AAUC18Survey,Marcolla22Survey,%
  ViandJattkeHithnawi21SoKCompilers,SoKFHEAccel24,VIPBench21,HEProfiler24%
}

\HEdefcat{hardware}{%
  DaiSunar15cuHE,NuFHE,Jung21CKKS100x,TensorFHE23HPCA,PhantomFHE23,%
  Shen23CARM,Badawi18GPUFV,%
  RoyTuran19HPCA,HEAX20,TuranRoy20HEAWS,HEXLFPGA21,FAB23HPCA,%
  Poseidon23HPCA,FxHENN23HPCA,FAST25FPGA,MAD23,%
  F121MICRO,CraterLake22ISCA,BTS22ISCA,ARK22MICRO,SHARP23ISCA,%
  BASALISC22,SoKFHEAccel24,%
  CryptoPIM20,gupta2022memfhe,CiMHE19,gupta2023evaluating,Cabrera24HEPIM,%
  zhou2025fhemem,HMCFHE24,%
  CIPHERMATCH25,FlashCosmos22,ParaBit21,AresFlash24,TawoseHE23%
}

\HEdefcat{hwall}{%
  DaiSunar15cuHE,NuFHE,Jung21CKKS100x,TensorFHE23HPCA,PhantomFHE23,%
  Shen23CARM,Badawi18GPUFV,%
  RoyTuran19HPCA,HEAX20,TuranRoy20HEAWS,HEXLFPGA21,FAB23HPCA,%
  Poseidon23HPCA,FxHENN23HPCA,FAST25FPGA,MAD23,%
  F121MICRO,CraterLake22ISCA,BTS22ISCA,ARK22MICRO,SHARP23ISCA,%
  BASALISC22,SoKFHEAccel24,%
  CryptoPIM20,gupta2022memfhe,CiMHE19,gupta2023evaluating,Cabrera24HEPIM,%
  zhou2025fhemem,HMCFHE24,%
  CIPHERMATCH25,FlashCosmos22,ParaBit21,AresFlash24,TawoseHE23,%
  PipeZK21ISCA,SecureML17,ABY3_18%
}

\HEdefcat{allhe}{%
  RAD78PrivacyHomo,Regev09LWE,LPR10RLWE,LPR13RLWEjournal,Gentry09FHE,%
  Gentry09Thesis,BV11LWEFHE,BGV12LeveledFHE,BGV14LeveledFHEjournal,%
  SV14FHESIMD,GSW13,HS20HElibDesign,DGHV10,Brakerski12ScaleInv,%
  GH11Implementing,AAUC18Survey,Peikert16Decade,%
  BV11RLWE,FV12BFV,CKKS17,CHKKS18Bootstrap,CHKKS18FullRNS,%
  CCS19ImprovedBootstrap,DM15FHEW,CGGI16FasterBootstrap,CGGI20TFHE,%
  HS15HElibBoot,MP21BootstrappingFHEW,%
  OpenFHE22,SEAL,ChenLainePlayer17SEAL,HElibRepo,TFHELib,Concrete20,%
  TFHErs,Lattigo,TenSEAL21,IbarrondoViand21Pyfhel,PhantomFHE23,%
  PALISADE,HEIR,%
  CHET19PLDI,EVA20PLDI,nGraphHE19,nGraphHE2_19,Cingulata15,%
  Porcupine21PLDI,HECO23,Fhelipe24PLDI,ViandJattkeHithnawi21SoKCompilers,%
  Marble18,%
  gilad2016cryptonets,LoLa19,Gazelle18,Delphi20,ReagenCheetah21,HuangCheetah22,%
  LeeMPCNN22,ConcreteML,%
  HE3DB23,ArcEDB24,HEDA22,CIPHERMATCH25,SealPIR18,OnionPIR21,Spiral22,%
  SimplePIR23,ChenLaineRindal17PSI,%
  KimEtAl21Genotype,LauterNaehrigVaikuntanathan11,GraepelLauterNaehrig12,%
  CryptoGCN22,SavPoseidon21,%
  HS14AlgorithmsHElib,HS18FasterLin,Bossuat21EUROCRYPT,%
  IliashenkoZucca21PoPETs,HalevPolyakovShoup19,KimPolyakovZucca21,%
  ChillottiCGGI17Packed,%
  DaiSunar15cuHE,NuFHE,Jung21CKKS100x,TensorFHE23HPCA,Shen23CARM,%
  Badawi18GPUFV,%
  RoyTuran19HPCA,HEAX20,TuranRoy20HEAWS,HEXLFPGA21,FAB23HPCA,%
  Poseidon23HPCA,FxHENN23HPCA,FAST25FPGA,MAD23,%
  F121MICRO,CraterLake22ISCA,BTS22ISCA,ARK22MICRO,SHARP23ISCA,%
  BASALISC22,SoKFHEAccel24,%
  CryptoPIM20,gupta2022memfhe,CiMHE19,GuptaIISWC23HEPIM,Cabrera24HEPIM,%
  zhou2025fhemem,HMCFHE24,%
  FlashCosmos22,ParaBit21,AresFlash24,TawoseHE23,%
  PipeZK21ISCA,SecureML17,ABY3_18,%
  HEStandard18,Marcolla22Survey,VIPBench21,HEProfiler24%
}

\makeatother

\def\BibTeX{{\rm B\kern-.05em{\sc i\kern-.025em b}\kern-.08em
    T\kern-.1667em\lower.7ex\hbox{E}\kern-.125emX}}

\title{\LARGE{HE-PIM: Demystifying Homomorphic Operations \\ on a Real-world Processing-in-Memory System}}

\author{
    *Harshita Gupta$^1$ \hspace{0.5em}
    *Mayank Kabra$^1$ \hspace{0.5em}
    Jaewoo Park$^1$ \hspace{0.5em}
    Priyam Mehta$^2$ \vspace{0.2em} \\
    Phillip Widdowson$^1$ \hspace{0.5em}
    Tathagata Barik$^{2,3}$ \hspace{0.5em}
    Nisa Bostanci$^1$ \hspace{0.5em}
    Konstantinos Kanellopoulos$^1$ \vspace{0.2em} \\
    Juan G\'omez-Luna$^4$ \hspace{0.5em}
    Antonio J. Pe\~na$^2$ \hspace{0.5em}
    Mohammad Sadrosadati$^1$ \hspace{0.5em}
    Onur Mutlu$^1$ \vspace{0.5em} \\
    \normalsize{
        $^1$ETH Zürich \hspace{0.5em}
        $^2$Barcelona Supercomputing Center \hspace{0.5em}
        $^3$Universitat Polit\`ecnica de Catalunya \hspace{0.5em}
        $^4$NVIDIA
    }
}

\begin{document}

\bstctlcite{IEEEexample:BSTcontrol}

\sloppypar



\maketitle

\def\thefootnote{}\footnotetext{*Harshita Gupta and Mayank Kabra are co-primary authors.}\def\thefootnote{\arabic{footnote}}

\thispagestyle{firstpage}

\begin{abstract}
Homomorphic encryption (HE) enables computation over encrypted data without decryption, offering strong privacy guarantees for untrusted computing environments. However, its practical adoption remains limited by high computational complexity, large ciphertext sizes, and substantial data movement. Processor-centric architectures (e.g., CPUs, GPUs, and ASICs) face fundamental bottlenecks when executing HE workloads because ciphertexts are large, HE operations exhibit low data locality, and primitives such as relinearization and bootstrapping require repeated access to large auxiliary metadata. Processing-In-Memory (PIM) is a promising approach to mitigate this data-movement bottleneck by performing computation near or inside memory. Several prior works propose PIM-based implementation for homomorphic operations; however, they either do not consider \emph{real-world} PIM systems or focus on a limited subset of homomorphic operations.

Our goal is to comprehensively characterize homomorphic operations on a real-world, general-purpose PIM system. To this end, we implement a complete set of HE kernels, which are used to implement emerging applications (e.g., databases and machine learning) on the \emph{real-world} UPMEM PIM system, rigorously evaluate these kernels in terms of performance and scalability, compare against CPU and GPU baselines, and discuss implications for future PIM hardware.

Our results demonstrate four major findings. (1) HE-based applications expose different bottlenecks across different stages of execution: some kernels are compute-bound due to expensive modular arithmetic, while others are constrained by the large memory footprint of ciphertexts and intermediate data. In our baseline PIM system, these bottlenecks are exacerbated by the computational capability of each PIM core and the limited capacity of each PIM bank, leading to frequent data movement. (2) The dominant compute bottleneck in current PIM cores is the lack of native 64-bit modular integer multiplication, which is a key primitive in HE operations and subroutines. (3) The limited memory capacity of each PIM bank is a second major bottleneck, as HE workloads use large ciphertexts and their associated auxiliary metadata that do not fit in the current PIM bank, requiring data movement across PIM cores and banks to accommodate large ciphertexts and intermediate results. (4) Despite these limitations, PIM systems can be a viable alternative to state-of-the-art CPU and GPU systems for HE workloads when equipped with native modular multiplication and efficient inter-PIM data movement.
\end{abstract}

\section{Introduction}


Homomorphic encryption (HE)~\citecat{foundations} provides a strong cryptographic foundation for privacy-preserving computation by enabling computations on encrypted data without exposing the underlying plaintext to untrusted systems. This property makes HE attractive for secure applications such as machine learning inference~\citecat{ml} and encrypted database search~\citecat{databases}. Despite significant improvements in HE algorithms, HE remains computationally expensive in practice due to two major challenges.
First, HE significantly increases data size: a small
unencrypted vector (plaintext) 
(e.g., 0.5MB)
can expand to a large encrypted tensor (ciphertext) of more than 30MB~\cite{kabra2025ciphermatch, kim2025anaheim, gupta2022memfhe,samardzic2022craterlake}.
As a result, processor-centric architectures (e.g., CPUs, GPUs, and ASICs) frequently move multiple ciphertexts between the processor and off-chip main memory (DRAM)~\citecat{membottleneck} (\S\ref{subsec:he}). 
Second, modern privacy-preserving applications (e.g., secure neural network inference) involve complex homomorphic operations (e.g., multiplication, rotation, relinearization, and bootstrapping)
that frequently access large auxiliary data (e.g., relinearization or bootstrapping keys) (\S\ref{subsec:ckks}). This auxiliary data is often 10-100 MB in size~\cite{kim2022ark,samardzic2022craterlake}, do not fit in local caches, and are transferred repeatedly between memory and compute units, further exacerbating the data movement bottleneck. 
\emph{Processing-in-memory} (PIM)~\cite{mutlu2022modern,mutlu2023dac,ghose2019processing,mutlu2019enabling} is a promising approach to alleviating the data movement bottleneck by placing computational units inside or near memory. Recently, this memory-centric computing paradigm has gained traction in both academia and industry 
(e.g.,~\citecat{PIM});
and commercial PIM systems and prototypes have recently been developed by industry~\cite{upmemwebsite2024, upmemtechpaper2022,upmemproductsheet2022,devaux2019true,lee2021hardware,kwon202125,ke2021near,lee2022isscc,niu2022184qps} (\S\ref{subsec:pim}).




Several prior works (e.g.,~\citecat{pim}) explore the effectiveness of using PIM architectures to
fundamentally improve the performance and energy
efficiency of homomorphic computations.
However, none of these prior works provides a comprehensive evaluation of these homomorphic operations used for building modern applications on real-world general-purpose PIM architectures. To our knowledge, there is only one prior work~\cite{gupta2023evaluating} on evaluating homomorphic operations on a real-world PIM system, and a complementary recent work~\cite{barik2026long} that studies the performance analysis of long-integer NTT execution for HE.
Unfortunately, these works discuss only basic homomorphic operations~\cite{fan2012somewhat}, i.e., homomorphic addition and multiplication (which \emph{only} use element-wise operations), omitting the complex homomorphic subroutines (e.g., NTT, relinearization, bootstrapping) that are required to construct modern applications (\S\ref{sec:motivation}). 

\textbf{Our goal} in this work is to understand the capabilities and characteristics of these homomorphic operations used in
various modern applications on a real-world PIM system and understand their implications for PIM systems. 
We focus on one of the most common HE workload, i.e., neural network inference~\cite{gilad2016cryptonets,LoLa19}, which involves complex convolution layers and non-linear activation functions and requires all HE operations/subroutines used in fully homomorphic encryption (FHE)~\cite{gentry2011implementing, gentry2012fully, gentry2012better, smart2014fully, chatterjee2019fully}. Hence, neural network inference serves as a useful application for characterization. 

We develop an end-to-end implementation of core HE primitives/subroutines on a \emph{real-world} UPMEM PIM architecture 
to support FHE workflows. We choose the general-purpose UPMEM PIM system for our study~\cite{upmemwebsite2024,upmemtechpaper2022,upmemproductsheet2022,devaux2019true,gomez2021benchmarking,gomez2022benchmarking,falevoz2024energy}. Furthermore, the UPMEM system closely resembles other PIM systems (e.g.,~\cite{kim2021aquabolt,kwon2022system})  
and 
can provide insights for the design of future PIM systems~\cite{hyun2024pathfinding,jonatan2024scalability}.  
First, we map the core homomorphic operations (addition, multiplication, relinearization, bootstrapping) to fit the architectural constraints of the real-world PIM system. 
These
operations rely on a set of underlying computational primitives (e.g.,
modular arithmetic, number-theoretic transforms (NTT and INTT), matrix-vector multiplication, and element-wise arithmetic operations). We implement these primitives by leveraging the parallel execution model of the PIM architecture to maximize parallelism
and throughput (\S\ref{sec:implementation}).
Second, we characterize a range of HE
parameters used in
real-world applications and investigate the impact of different configurations on the performance 
and scalability of homomorphic operations on PIM systems (\S\ref{sec:methodology}).
We compare the UPMEM PIM system implementation to state-of-the-art CPU (AMD EPYC 7742 CPU~\cite{amdepyc2019}) and GPU (NVIDIA A100~\cite{nvidia2020a100}) baselines (\S\ref{sec:evaluation}). Third, we discuss the implications for future PIM systems and highlight the need to shift to an algorithm-hardware co-design perspective to support large-scale HE
applications running on a PIM system.

\textbf{Our results} demonstrate four key takeaways: (1) 
The performance of HE kernels on existing PIM systems is constrained by both the computational capabilities of each PIM unit and the memory capacity of each memory bank, leading to frequent data movement. This is because homomorphic operations/subroutines require complex arithmetic and have a substantial memory footprint (\S\ref{subsec:eval_nn}, \ref{sec:gpu}, \ref{sec:eval_end-nn}). (2) The primary computational bottleneck in current PIM cores arises from the lack of native 64-bit modular integer multiplication. We show that augmenting each PIM core with a 64-bit modular multiplier can improve the performance of existing homomorphic operations/subroutines by up to $2.3\times$ over the state-of-the-art UPMEM-PIM system. (3) The limited capacity of each PIM bank is another major limiting factor that necessitates frequent data movement between PIM cores. We show that supporting efficient data movement between PIM cores, along with a 64-bit modular multiplier, dramatically improves performance. Specifically, a 2048-core PIM system shows a potential headroom improvement of improving existing PIM systems by up to $426.1\times$ for convolution layer and $641.0\times$ for key switching operations on $32\times32$ image inputs, outperforming both CPU and GPU baselines (\S\ref{sec:discussion}). (4) PIM systems can be a viable alternative over state-of-the-art CPU and GPU systems for implementing HE workloads. This is supported by the fact that PIM systems can scale to thousands of cores for computation-dominated kernels and can be extended further. In comparison, GPU systems have a limited memory capacity of up to 80 GB per GPU and remain constrained by effective bandwidth (\S\ref{sec:final_discussion}).

This paper makes the following key contributions:
\squishlist
\item 
To our knowledge, this is the first end-to-end implementation of widely used homomorphic operations/subroutines on a \emph{real-world} Processing-In-Memory (PIM) system.
\item We perform a comprehensive characterization of various HE parameters and evaluate the performance of key homomorphic operations/subroutines required by a general HE-based application
on a state-of-the-art \emph{real-world} PIM system, and compare against CPU and GPU baselines.
\item We analyze key system-level trade-offs and provide practical design recommendations for future HE software stacks and PIM hardware architectures, highlighting the importance of algorithm-hardware co-design.
\squishend

\section{Background}
\label{sec:background}



\subsection{Homomorphic Encryption}
\label{subsec:he}
Homomorphic encryption (HE)~\cite{Gentry09FHE} enables computation directly on encrypted data and consists of four fundamental algorithms:
\emph{(i)} \textsf{KeyGen}, which generates a secret key (and optionally a public key and evaluation keys);
\emph{(ii)} \textsf{Enc}, which transforms a plaintext into a ciphertext by embedding it in a structured polynomial ring and adding a small random error (noise), ensuring semantic security;
\emph{(iii)} \textsf{Eval}, which performs algebraic operations (e.g., addition, multiplication) directly on ciphertexts, while also increasing the error (noise) associated with each input; and
\emph{(iv)} \textsf{Dec}, which uses the secret key to recover the plaintext from a ciphertext, as long as the accumulated noise remains below a decryption threshold.
For a function $f$ and message $m$, HE ensures: 
\(
\textsf{Dec}(\textsf{Eval}(f, \textsf{Enc}(m))) \approx f(m).
\)


\textbf{Fully Homomorphic Encryption (FHE)}~\cite{gentry2011implementing, gentry2012fully, gentry2012better, smart2014fully, chatterjee2019fully}
 extends the basic capabilities of HE by supporting unlimited additive and multiplicative operations over ciphertexts. 
Multiple FHE schemes have been proposed in the literature (e.g., BFV~\cite{Brakerski12ScaleInv}, BGV~\cite{BGV12LeveledFHE}, CKKS~\cite{cheon2017homomorphic}), each offering different trade-offs between exactness, performance, and noise management. In this work, we focus on the CKKS scheme, which supports approximate arithmetic over real or complex numbers and is well-suited for applications such as encrypted neural network inference.



\subsection{The CKKS Scheme}
\label{subsec:ckks}

CKKS~\cite{cheon2017homomorphic} supports efficient linear algebra operations by encoding 
real-valued vectors into the polynomial ring:
\(
R_Q = \mathbb{Z}_Q[X] / (X^N + 1),
\)
where $N$ is a power-of-two polynomial degree and $Q$ is a large ciphertext modulus. Each ciphertext consists of two polynomials in $R_Q$: \((c_0,c_1)\), with coefficients represented using the Residue Number System (RNS) \cite{bajard2016full}. In RNS form, each polynomial coefficient (modulo $Q$) is decomposed into a set of $L+1$ smaller primes, producing $L+1$ polynomials with smaller coefficients (referred to as \emph{limbs}) that enable fast modular arithmetic and parallelism across residues.
A key feature
of CKKS is its support for \emph{SIMD-style packing}, where $N$ real (or complex) values can be packed into a single ciphertext, enabling parallel element-wise operations across all slots. CKKS encodes a vector $m \in \mathbb{C}^{n}$ into a plaintext polynomial $[m] \in R_Q$ by applying the inverse of the canonical embedding (structurally analogous to an inverse FFT over roots of the cyclotomic polynomial). This maps the values into the polynomial domain and scales them by a large factor $\Delta$ to preserve precision during integer-based polynomial arithmetic. Homomorphic operations such as addition and multiplication are then performed on polynomials, corresponding to slot-wise operations on the encoded vector.

Table~\ref{tab:ckks_params} summarizes the HE notations used in this work.

\begin{table}[h]
\footnotesize
\centering
\begin{tabular}{ll}
\toprule
\textbf{Symbol} & \textbf{Description} \\
\midrule
$m$ & Cleartext vector of real or complex values \\
$[m]$ & Plaintext polynomial encoding $m$ \\
$\llbracket m \rrbracket$ & Ciphertext encrypting plaintext $[m]$ \\
$N$ & Polynomial degree\\
$Q$ & Ciphertext modulus, product of primes $q_i$ \\
$|Q|$ & Bit-length of Q \\
$q_i$ & i-th RNS prime modulus composing $Q$ \\
$L$ & Maximum multiplicative level (modulus chain length) \\
$\ell$ & Current level of a ciphertext ($0 \leq \ell \leq L$) \\
$\Delta$ & Scaling factor used in encoding \\
$L_{\text{boot}}$ & Levels consumed during bootstrapping \\
$L_{\text{eff}}$ & Effective usable depth after bootstrapping \\
\bottomrule
\end{tabular}
\caption{Relevant CKKS notation.}
\label{tab:ckks_params}
\end{table}



\subsubsection{Homomorphic Operations}
\label{sec:hom_ops}
We describe the homomorphic operations supported by CKKS: addition, multiplication, and rotation, as well as primitives such as NTT/INTT. 

\noindent \textbf{Addition.}
CKKS supports
element-wise addition between a ciphertext and a plaintext (\textsf{PAdd}: $[a] + \llbracket b \rrbracket$) or between two ciphertexts (\textsf{CAdd}: $\llbracket a \rrbracket + \llbracket b \rrbracket$). In both cases, the operands must reside at the same multiplicative level and have identical scaling factors. The result is a ciphertext $\llbracket c \rrbracket_{\text{add}}$ with the same level and scale. Functionally, this represents a SIMD (element-wise) addition of the underlying cleartext vectors:
\(
\texttt{decode}(\texttt{decrypt}(\llbracket c \rrbracket_{\text{add}})) \approx a \oplus b.
\)
Addition does not consume any multiplicative levels (L).

\noindent \textbf{Multiplication.}
CKKS supports multiplication between a ciphertext and a plaintext (\textsf{PMul}: $[a] \cdot \llbracket b \rrbracket$) or between two ciphertexts (\textsf{CMul}: $\llbracket a \rrbracket \cdot \llbracket b \rrbracket$). CMul increases the ciphertext’s scale from $\Delta$ to $\Delta^2$ and produces a three-component ciphertext $(d_0, d_1, d_2)$. To manage scale growth and maintain correctness, a \texttt{rescale} step is applied:
\(
\texttt{Rescale}(\llbracket c \rrbracket): \Delta^2 \rightarrow \Delta,
\)
which drops the last modulus in the RNS chain and reduces the multiplicative level by one. CMul also requires \texttt{relinearization}, which eliminates the $d_2$ term via key switching (\S\ref{sec:subroutines}). The resulting ciphertext $\llbracket c \rrbracket_{\text{mul}}$ is returned at level $\ell - 1$ with restored scaling factor $\Delta$.

\noindent \textbf{Rotation/Automorphism.}
CKKS supports cyclic rotations (\textsf{HRot}) of cleartext slots by a user-specified offset $k$:
\(
\llbracket c \rrbracket' = \texttt{Rotate}(\llbracket c \rrbracket, k).
\)
Rotations are implemented via Galois automorphisms
followed by key switching using 
rotation keys. Rotations preserve the level and scale of the ciphertext
without consuming multiplicative depth. This operation is essential for vector manipulations, such as matrix multiplication, convolution, and reduction.

\noindent \textbf{NTT and INTT.} Since homomorphic encryption requires frequent polynomial multiplication (naively $O(n^2)$), homomorphic encryption libraries~\citecat{libraries} store the ciphertext in a number theoretic transform (NTT) form. 
For a given polynomial $a = a_0 + a_1x + a_2x^2 + \ldots + a_{N-1}x^{N-1}$, we represent its coefficients as a vector $\mathbf{a} = (a_0, a_1, \ldots, a_{N-1})$. The polynomial multiplication $\mathbf{a} \ast \mathbf{b}$ can be computed as follows:
\(
    \mathbf{a} \ast \mathbf{b} = NTT^{-1}(NTT(\mathbf{a}) \odot NTT(\mathbf{b}))
\)
where $\odot$ represents the element-wise product and $NTT^{-1}$ is the inverse NTT.
Because the NTT shares characteristics with the finite-field equivalent of the DFT, it can be computed using the Cooley-Tukey algorithm~\cite{cooley1965algorithm,satriawan2024complete}. An $N$-point NTT can be computed in $\log_2 N$ stages, each of which involves $N/2$ butterfly operations, reducing computational complexity to $O(n\log n)$. 

\subsubsection{Homomorphic Subroutines}
\label{sec:subroutines}
Based on these homomorphic operations and NTT/INTT primitives, we describe two homomorphic subroutines: key switching and bootstrapping.

\noindent \textbf{Key Switching.} When a ciphertext is rotated, or its structure is changed (e.g., after CMul generates the $d_2$ term), it cannot be decrypted under the same secret key. Hence, the ciphertext must be transformed back to the original key using the key-switching mechanism. This process requires three steps. (1) \textsf{ModUp}, where the ciphertext modulus $Q$ is increased to an extended modulus $P \cdot Q$ through an inverse NTT (on the ciphertext), basis extension (BConv) (which becomes a matrix multiplication operation after performing INTT on the ciphertext), and a forward NTT (to return to the original slot domain). (2) \textsf{EvalKey}, where a precomputed evaluation key is multiplied with the changed component in the ciphertext using PMul. (3) Finally, \textsf{ModReduce} removes the extended $P$ basis, returning the ciphertext to the original $Q$ domain.

\noindent \textbf{Bootstrapping.}
When a ciphertext reaches level zero, further 
homomorphic 
operations become infeasible due to excessive noise. CKKS supports bootstrapping \cite{ckksbootstrapping,han2020better,lee2021high} to refresh such ciphertexts by homomorphically evaluating an approximate identity function over the encrypted data. The process requires five steps. (1) \textsf{ModUp}, where the ciphertext modulus $Q$ is increased to an extended modulus $P \cdot Q$, as in key switching. (2) The \textsf{Slots-to-coefficient (S2C)} transformation is applied using an inverse discrete Fourier transform (IDFT), which extracts the SIMD-encoded slots (of the plaintext hidden under the ciphertext) to the coefficient domain. (3) The \textsf{EvalMod} operation is performed in the slot domain, where a low-degree Chebyshev polynomial is evaluated to approximate the identity function~\cite{kim2022approximate}. 
(4) \textsf{Coefficient-to-slot (C2S)} transformation is applied, which performs a discrete Fourier transform (DFT) to bring the result back to the slot domain for the next homomorphic computations. (5) Finally, \textsf{ModDown} removes the extended $P$ basis, returning the ciphertext to the original $Q$ domain. Bootstrapping consumes a fixed number of multiplicative levels, $L_{\text{boot}}$, yielding an effective usable depth $L_{\text{eff}} = L - L_{\text{boot}}$. Bootstrapping is essential for supporting arbitrarily deep homomorphic computations; however, it is computationally expensive.



%

\subsection{Processing-In-Memory}
\label{subsec:pim}
Processing-in-memory (PIM)~\citecat{PIM}
is a computing paradigm that aims to address the data movement bottleneck that occurs when data must be moved between memory 
and compute units in traditional processor-centric systems~\citecat{membottleneck}. This bottleneck is caused by a substantial performance gap
between processors and memory, often referred to as the memory wall. 
PIM architectures are becoming a reality, with commercialized systems like the UPMEM PIM architecture~\cite{devaux2019true,gomez2021benchmarking,gomez2022benchmarking,hyun2024pathfinding} and prototyped systems like HBM-PIM~\cite{kim2021aquabolt, kim2022aquabolt}, AxDIMM~\cite{lee2022improving, ke2021near}, and HB-PNM~\cite{niu2022184qps} showing significant performance benefits for data-intensive applications.

\section{Motivation for using PIM}
\label{sec:motivation}

Prior architectural studies have largely adopted compute-centric paradigms, optimizing arithmetic throughput for core FHE primitives such as NTTs, polynomial multiplication, and bootstrapping. On CPUs, software-level optimizations such as vectorization (e.g., AVX2/AVX-512), multithreading, cache tuning, and RNS-based arithmetic optimizations improve computational performance~\cite{hexl,sealcrypto}. However, these approaches are fundamentally constrained by limited memory bandwidth, data transfer between deep cache hierarchies, and insufficient parallelism for FHE’s batched SIMD structure (\S\ref{subsec:he}). As a result, even optimized CPU implementations exhibit high latency in end-to-end FHE workloads (e.g., neural network inference). GPUs offer high arithmetic throughput and wide SIMD parallelism, accelerating basic FHE primitives~\cite{dai2016cuhe,cuFHE,kim2022ark}. However, limited on-chip memory (e.g., a 40 MB L2 cache) and reliance on off-chip DRAM ($<$2 TB/s) lead to frequent data movement, especially when FHE ciphertexts (e.g., a single ciphertext of 40 MB~\cite{kim2022ark}) and evaluation keys (e.g., more than 120 MB~\cite{kim2022ark}) exceed available on-chip capacity. FPGAs enable custom datapaths for FHE operations but face challenges in resource reuse and bandwidth efficiency~\cite{cao2015optimised,cao2013accelerating,roy2018hepcloud,roy2019fpga}, limiting their scalability. ASICs achieve significant speedups through tightly integrated designs by deploying 512 MB of on-chip cache~\cite{feldmann2021f1,samardzic2022craterlake,kim2022bts,kim2023sharp}. However, their designs are rigid, lack adaptability to evolving FHE schemes, and incur high design costs (e.g., 512 MB of on-chip cache with 2048 custom-logic SIMD lanes).

\begin{figure}[h]
\includegraphics[width=1.0\linewidth]{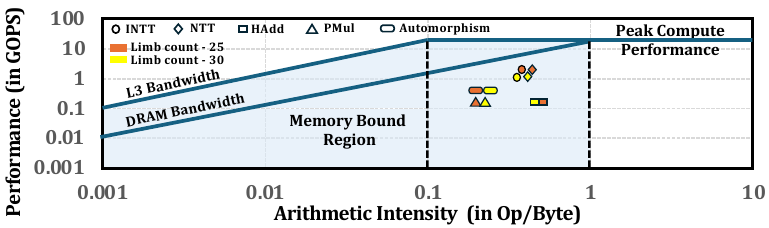}
\caption{Roofline Model of the homomorphic operations (HAdd, PMul, NTT, INTT, and automorphism).}
\label{fig:roofline}
\end{figure}

Hence, prior compute-centric designs largely focus on improving the performance of homomorphic operations by optimizing under the assumption that improvements to compute units are the primary path to better performance. To understand the performance tradeoffs of HE operations, we employ a roofline model~\cite{williams2009roofline} to visualize the extent to which homomorphic operations are constrained by memory bandwidth and computational limits. Figure~\ref{fig:roofline} shows the roofline model on an AMD EPYC 7742 CPU (\S\ref{sec:methodology}). We evaluate four key primitives at two different limb counts, 25 and 30\footnote{This means that a large ciphertext modulus (\(Q\)) is broken into an RNS chain (\(q_i\)) of 25 and 30 small prime moduli \(Q = q_1*q_2*\ldots*q_{25/30}\) (\S\ref{subsec:ckks}).}:
(1) element-wise addition, (2) element-wise multiplication, (3) NTT/INTT, and (4) automorphism.
We observe that all of these kernels are fundamentally limited by memory bandwidth.\footnote{The shaded area at the intersection of DRAM bandwidth and the peak compute performance roof is the memory-bound region in the roofline plot.} This data movement bottleneck is further exacerbated when using more compute power (e.g., GPUs) with a small cache to buffer intermediate data. Hence, performance of homomorphic operations can be improved by reducing data movement bottlenecks, making them well-suited for PIM systems~\cite{gomez2022benchmarking,mutlu2022modern,oliveira2021pimbench,draper2002architecture,devaux2019true,ghose2019processing,falevoz2024energy}. 

Prior HE-PIM designs~\cite{gupta2022memfhe, reis2020computing, gupta2021accelerating,zhou2025fhemem} primarily improve HE performance by adding custom logic near memory. While these designs demonstrate the potential of near-memory execution, they leave open how much of this opportunity can be realized by existing programmable PIM systems and what targeted hardware improvements are required to close the gap. Rather than assuming a custom PIM substrate, we begin with UPMEM, a commercially available PIM architecture~\cite{upmem2018}, quantify the bottlenecks that limit its effectiveness for HE, and model the architectural enhancements needed to make such programmable PIM a viable substrate for HE acceleration.
To our knowledge, only one prior work by Gupta et al.~\cite{gupta2023evaluating} investigates the potential
of a real-world UPMEM PIM system to accelerate
homomorphic operations. Unfortunately, their work only evaluates two basic homomorphic operations ($\mathsf{HAdd}$ and $\mathsf{PMul}$) (\S\ref{subsec:he}), which use element-wise multiplication and addition kernels on the UPMEM PIM system (\S\ref{sec:elem_ops_impl}). These primitives are limited and primarily support only simple applications (e.g., linear or logistic regression). A general HE application involves all the homomorphic operations and subroutines described in \S\ref{subsec:he}, combined in complex ways.
Recent work on long-integer NTT execution on UPMEM-PIM~\cite{barik2026long} further highlights the relevance of real PIM hardware for FHE kernels, but it does not evaluate the full HE operation pipeline needed by end-to-end applications.
Notably, the key HE operations in Figure~\ref{fig:roofline} lie near the memory-bound region, implying that higher bandwidth can shift them toward the compute-bound regime.
Hence, an in-depth architectural analysis of a PIM system for all homomorphic operations is required to understand the characteristics of homomorphic operations from a memory-centric computing perspective, thereby enabling us to design and implement better PIM systems.

\textbf{Our goal} is to provide a comprehensive experimental characterization of homomorphic operations used in HE applications on a real-world PIM architecture to understand their implications for PIM systems. To do so, we develop an end-to-end implementation of core HE operations and subroutines on a real-world UPMEM PIM system to support FHE workflows.

\section{Methodology}
\label{sec:methodology}

To understand the performance impact on a PIM system, we implement the homomorphic operations and subroutines on a real-world UPMEM PIM system for our evaluation. We compare the performance of these implementations used across various parameters in an end-to-end privacy-preserving neural network on the PIM system with state-of-the-art CPU and GPU systems.

\begin{figure}[h]
    \centering
    \includegraphics[width=0.9\linewidth]{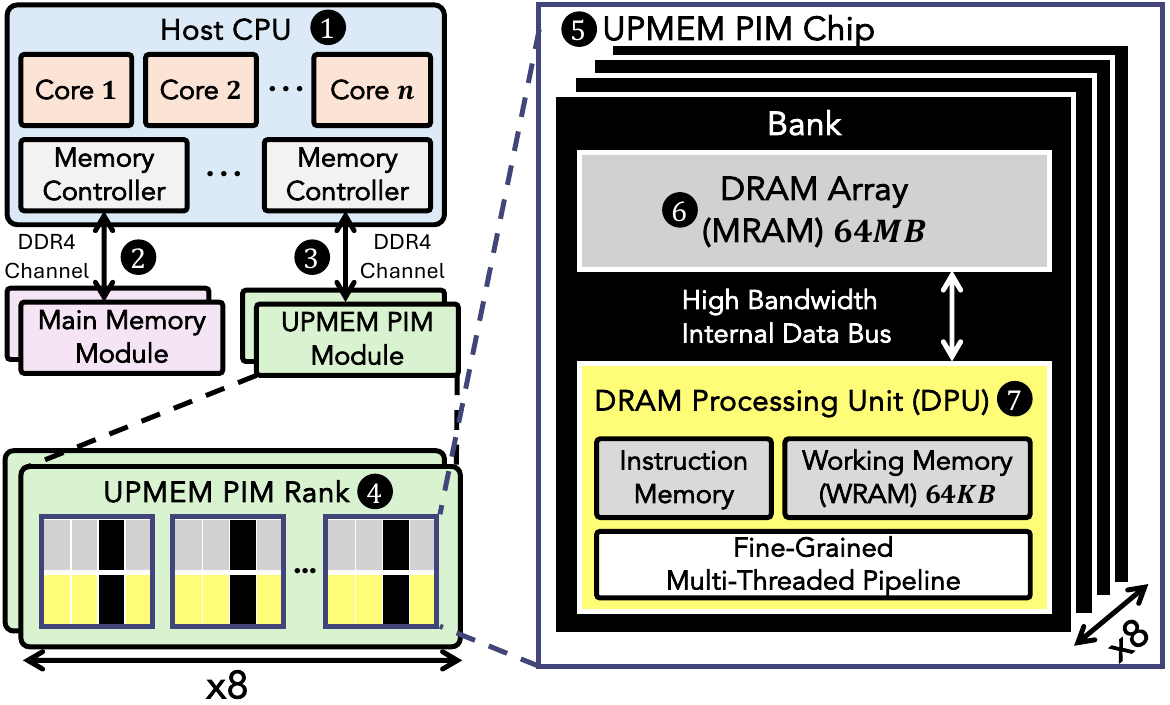}
    \caption{High-level system organization of the UPMEM PIM system and the hardware architecture of a UPMEM PIM chip.}
    \label{fig:upmem}
\end{figure}

\noindent \textbf{The UPMEM PIM Architecture.}
\label{sec:UPMEM_PIM_arch} Fig.~\ref{fig:upmem} shows the high-level system organization of a UPMEM PIM 
system and the hardware architecture of a UPMEM PIM chip. The system includes a regular host CPU~\circled{1}, conventional main memory modules~\circled{2}, and UPMEM PIM memory modules~\circled{3}. Each UPMEM PIM memory module contains two \emph{ranks}~\circled{4}. Each rank has 8 UPMEM PIM \emph{chips}~\circled{5}. Inside each chip, there are 8 \emph{banks}. Each bank contains 1) a 64MB DRAM array called MRAM~\circled{6}, and 2) a \emph{general-purpose} \emph{DRAM Processing Unit} (DPU)~\circled{7} which is our PIM core.

The MRAM implements a standard JEDEC DDR4 DRAM interface accessible to the host CPU. The DPU has an SRAM Instruction Memory, a 64KB SRAM Working Memory (WRAM), and an in-order, fine-grained, multi-threaded pipeline with 11 stages that supports 24 hardware threads. It implements a 32-bit RISC-based Instruction Set Architecture (ISA) with native support for 32-bit integer addition/subtraction and 8-bit integer multiplication. Other more complex arithmetic operations (e.g., integer divisions and floating-point operations) are emulated through software. The DPU does not have a cache but uses the WRAM as a scratchpad memory~\cite{upmemproductsheet2022,upmemtechpaper2022}. Each PIM core has \emph{exclusive} access to its MRAM (with respect to other DPUs) through a high-bandwidth (up to 0.7GB/s per DPU) internal data bus~\cite{gomez2022benchmarking, gomez2022machine}. There are no direct communication channels among DPUs within the UPMEM PIM chip. All inter-DPU communication is handled by the host CPU (i.e., the host CPU first gathers data from the DPUs' MRAM into the system's main memory and then distributes it back to the DPUs' MRAM).

\noindent\textbf{PIM Programming and Execution Model.} PIM programs are written in the C programming language with the UPMEM SDK~\cite{upmem_sdk} and runtime libraries. The execution model of a PIM core is based on the \emph{Single-Program Multiple-Data} (SPMD) paradigm. Each PIM core runs multiple (up to 24) software threads, called \emph{tasklets}, which execute the same code but operate on different data. Each tasklet has its own control flow, independent from other tasklets. Tasklets are assigned to DPUs \emph{statically} by the programmer during \emph{compile-time}. Tasklets assigned to the same DPU share MRAM and WRAM~\cite{gomezluna2021prim, gomez2022experimental}. 

\subsection{System Configurations.} 
\label{sec:system_config}
\textbf{Performance Measurements.} Table~\ref{tab:dpu_system-configuration} shows the system configurations used in our experiments. 

\begin{table}[h]
\centering
\caption{Evaluated System Configurations.}
\resizebox{\linewidth}{!}{
\begin{tabular}{ll}
\hline
\multicolumn{2}{c}{\textbf{CPU Baseline System}} \\
\hline

\textbf{Processor}                                                   & \begin{tabular}[c]{@{}l@{}} AMD EPYC 7742 64-core processor @ 2.25GHz\end{tabular}  \\ \hline
\textbf{Main Memory}                                                        & \begin{tabular}[c]{@{}l@{}}\SI{1}{\tera\byte} total capacity\\32$\times{}$\SI{32}{\giga\byte} DDR4 (RDIMMs)\end{tabular}  \\ \hline
\hline
\multicolumn{2}{c}{\textbf{GPU Baseline System}} \\

\hline
\textbf{Processor}                                                   & \begin{tabular}[c]{@{}l@{}} Intel Xeon Gold 5118 12-core processor @ 2.30GHz\end{tabular}  \\ \hline
\textbf{Main Memory}                                                        & \begin{tabular}[c]{@{}l@{}}\SI{512}{\giga\byte} total capacity\\16$\times{}$\SI{32}{\giga\byte} DDR4 (RDIMMs)\end{tabular}  \\ \hline
\textbf{GPU}                                                        & \begin{tabular}[c]{@{}l@{}} 1$\times{}$ NVIDIA A100 (PCIe, \SI{80}{\giga\byte})\end{tabular}  \\ \hline

\hline
\hline

\multicolumn{2}{c}{\textbf{UPMEM PIM System}} \\

\hline
\textbf{Processor}                                                   & \begin{tabular}[c]{@{}l@{}} Intel Xeon Silver 4215 8-core processor @ 2.50GHz\end{tabular}  \\ \hline
\textbf{Main Memory}                                                        & \begin{tabular}[c]{@{}l@{}}\SI{256}{\giga\byte} total capacity\\4$\times{}$\SI{64}{\giga\byte} DDR4 (RDIMMs)\end{tabular}  \\ \hline
\begin{tabular}[c]{@{}l@{}}\textbf{PIM-Enabled}\\\textbf{Memory}\end{tabular} & \begin{tabular}[c]{@{}l@{}} \SI{160}{\giga\byte} total capacity, 20$\times{}$\SI{8}{\giga\byte} UPMEM PIM modules,\\2560 DPUs, 2 ranks per module, 8 chips per rank, 8 DPUs per chip\\\SI{350}{\mega\hertz} DPU clock frequency\end{tabular}   \\ \hline
\hline

\end{tabular}
}
\label{tab:dpu_system-configuration}
\end{table}

\textbf{Baseline CPU and GPU Implementation:} We use the CPU baseline system~\cite{amdepyc2019} with AMD EPYC 7742 64-core CPUs, and the GPU baseline system~\cite{nvidia2020a100} with an NVIDIA A100 GPU, on which we evaluate homomorphic operations. We implement our CPU and GPU baselines using the NEXUS framework~\cite{cryptoeprint:2024/136}, which extends Microsoft SEAL~\cite{sealcrypto} with GPU support and provides both CPU and GPU implementations of homomorphic operations. We implement all homomorphic operations and subroutines (\S\ref{sec:he-pim-subroutines}) and compare them individually against our UPMEM PIM implementation.\footnote{For a fair comparison, we do not use a cluster of GPUs for our baseline because the UPMEM PIM system is a single-server node. We leave the multi-GPU system comparison for future work.} 

\textbf{UPMEM-PIM System Configuration and Implementation:} 
The UPMEM PIM \cite{gomez2022benchmarking,nider2021case,devaux2019true}
system consists of a host CPU with standard main memory and PIM-enabled memory modules known as UPMEM DPU modules. Each module comprises two ranks, each containing 64 DRAM Processing Units (DPUs). 
Each DPU has exclusive access to 24~KB instruction memory (IRAM), 64~KB scratchpad memory (WRAM), and a 64~MB DRAM bank (MRAM), as shown in Figure \ref{fig:upmem}. The DPUs are multithreaded, in-order, 32-bit Reduced Instruction Set Computing (RISC) cores that can operate at up to 500 MHz. Each DPU has 24 hardware threads, each with 24 32-bit general-purpose registers. A DPU can perform single-cycle 8-bit multiplication. However, multiplication operations on 32-bit and 64-bit integers, and on 32-bit and 64-bit floats, are not supported by the current ISA and hardware; hence, they require longer software routines to execute. The Single Program Multiple Data (SPMD) programming paradigm is used on DPUs, where multiple software threads, called tasklets, execute the same code on different data. CPU--PIM and PIM--CPU data transfers can be performed in parallel across multiple DRAM banks. However, transfer sizes to and from all DRAM banks must be the same. With this architecture and its programming model, we describe our implementation of HE operations and subroutines in \S\ref{sec:implementation}.

\subsection{Workloads}
\label{sec:he_nn_impl}

We characterize a functionally complete set of HE kernels: any computation over encrypted data is decomposed into three homomorphic operations: addition ($\mathsf{HAdd}$, $\mathsf{PAdd}$), multiplication ($\mathsf{PMul}$, $\mathsf{CMul}$), and rotation ($\mathsf{HRot}$), together with two essential maintenance subroutines: relinearization (dominated by $\mathsf{BConv}$ and $\mathsf{PMul}$) and bootstrapping (dominated by $\mathsf{EvalMod}$) (\S\ref{sec:hom_ops}, \S\ref{sec:subroutines}). Based on these operations and kernels, we implement an end-to-end HE-based neural network as described below.


\subsubsection{HE-based Neural Network}

\begin{figure}[h]
    \centering
    \includegraphics[width=1\linewidth]{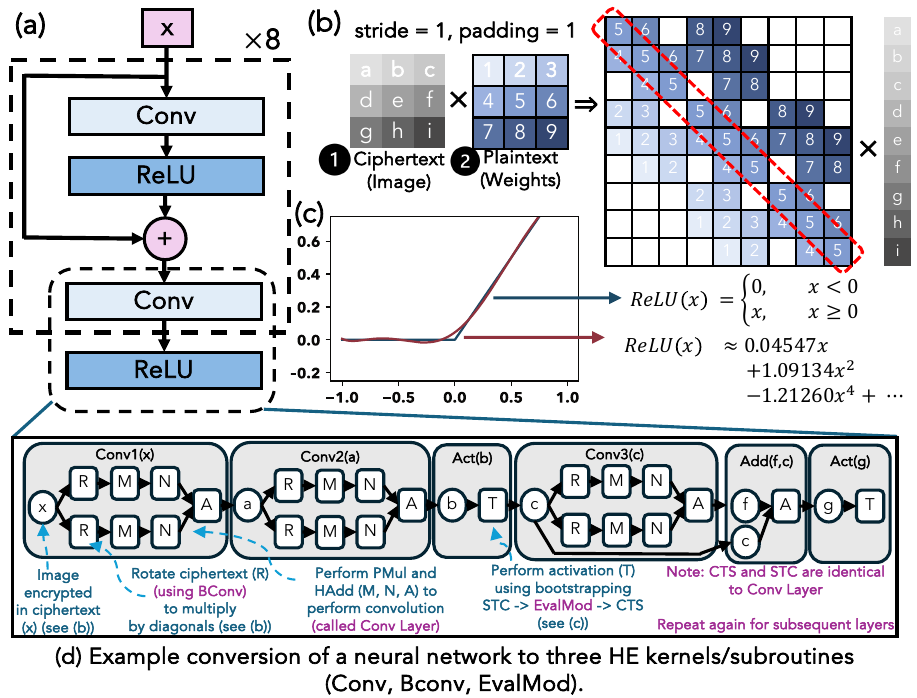}
    \caption{Overview of HE-based Neural Network Inference.}
    \label{fig:nn-inference}
\end{figure}

We assume that input images reside in the ciphertext domain, whereas model weights and layer parameters remain in plaintext form.\footnote{This threat model~\cite{kim2022ark,kim2022bts,samardzic2022craterlake,feldmann2021f1} ensures client privacy and preserves server secrecy by keeping model parameters locally on the server.} 
Figure~\hyperref[fig:nn-inference]{3(a)} illustrates the overall flow of HE-based neural network inference, which is expressed by decomposing each layer into a sequence of primitive HE operations and subroutines.
\textbf{Convolutional layer ($\mathsf{Conv}$)} adopts an encoding strategy that maps convolution to a homomorphic matrix–vector product. The input image is flattened and packed into the SIMD slots of a ciphertext, while the convolution kernel is encoded as a Toeplitz matrix of dimension
$T = (I-K+1) \times I$,
where each diagonal corresponds to a shift-and-weight operation applied to the input. Figure~\hyperref[fig:nn-inference]{3(b)} illustrates the use of primitive homomorphic operations as
\(\mathsf{PMul}\!\left(\mathsf{HRot}([\![t]\!],\, i \cdot 2^{k}\!\cdot s),\, P_{s,i}\right)\),
where \textsf{HRot} rotates the ciphertext~\circled{1} by the diagonal’s shift offset and \textsf{PMul} multiplies the rotated ciphertext with the corresponding plaintext weight vector \(P_{s,i}\)~\circled{2}. To reduce the cost of repeated rotations and plaintext multiplications, we employ the Baby-Step Giant-Step (BSGS) algorithm~\cite{halevi2012algorithms}, which restructures the Toeplitz matrix into a two-phase schedule that balances rotation count against multiplicative depth. BSGS decomposes the $T$ diagonals into baby steps of size $t_1$ and giant steps of size $t_2$ (where $t_1 \cdot t_2 = T$), lowering the number of required ciphertext rotations from $O(T)$ to $O(t_1 + t_2)$, which is minimized at $O(\sqrt{T})$ when $t_1 = t_2 = \sqrt{T}$. 


%

\textbf{Nonlinear activation functions} pose a fundamental challenge in HE, as ciphertexts cannot be compared or evaluated through non-polynomial functions. HE addresses this limitation by approximating common activations (e.g., ReLU, sigmoid) using low-degree Chebyshev polynomials~\cite{cheon2019polynomial,gilad2016cryptonets}, which can be efficiently evaluated homomorphically. Figure~\hyperref[fig:nn-inference]{3(c)} shows an approximation of the ReLU function using a Chebyshev approximation. However, even low-degree polynomial activations introduce significant multiplicative depth (consuming up to 16 multiplicative levels), particularly when composed across multiple layers. To mitigate this, prior works (e.g.,~\cite{lu2021pegasus,li2024evalcomp,alexandru2025general}) propose fusing the evaluation of the activation polynomial with the bootstrapping procedure, leveraging the fact that CKKS bootstrapping already involves homomorphic polynomial evaluation ($\mathsf{EvalMod}$, \S\ref{sec:subroutines}). By embedding the activation function within the bootstrapping pipeline, this fusion amortizes the depth and latency cost across multiple operations.

Figure~\hyperref[fig:nn-inference]{3(d)} shows that HE-based NN inference is decomposed into three dominant kernels: $\mathsf{Conv}$ (encrypted linear layers via rotations, ciphertext–plaintext multiplication ($\mathsf{PMul}$), and ciphertext–ciphertext multiplication ($\mathsf{CMul}$)), $\mathsf{BConv}$ (basis conversion for modulus/RNS switching), and $\mathsf{EvalMod}$ (modulus/noise management). 
 Thus, characterizing $\mathsf{Conv}$, $\mathsf{BConv}$, and $\mathsf{EvalMod}$ suffices to capture both individual operations and kernels required for end-to-end NN inference performance.

\section{Implementation}
\label{sec:implementation}

In this section, we describe our implementation details in three key steps. First, we analyze the inherent parallelism in CKKS ciphertexts, which informs our strategy for data mapping and workload distribution across PIM compute units. 
Second, we detail the simple element-wise operations and ciphertext multiplications (using NTT implementation). 
Third, we detail the key homomorphic primitives and subroutines required for HE-based neural network inference and describe how they are implemented under the resource and execution constraints of a real-world PIM system. 
The architectural characteristics of PIM guide our algorithmic choices and data layout strategies. Note that we target UPMEM as the only publicly available PIM platform~\cite{upmem2018,upmemproductsheet2022}; however, our implementation follows architecture-agnostic PIM principles (bank-local memory and massive parallelism) and thus generalizes to other PIM designs (e.g., Aquabolt~\cite{kim2021aquabolt}).

\subsection{Parallelism in HE Ciphertexts}
\label{sec:parallelism}
CKKS-based HE schemes expose three inherent levels of parallelism that enable efficient execution on architectures with many parallel compute units. First, ciphertext-level parallelism allows independent evaluation across multiple ciphertexts, which is especially beneficial in batched workloads where each ciphertext encodes a distinct portion of the input. Second, limb-level parallelism arises from the RNS representation of ciphertext coefficients, where arithmetic over independent residues can proceed in parallel within a single ciphertext. Third, CKKS’s slot-based encoding (canonical embedding) packs multiple values into independently operable SIMD slots, enabling vectorized operations inside each limb. Hence, parallelism across ciphertexts, RNS limbs, and SIMD slots forms a hierarchical execution structure well-suited to PIM architectures for efficiently scaling HE workloads.


\subsection{Homomorphic Operations on PIM}
\label{sec:he-pim-ops}

\subsubsection{Element-wise Operations}
\label{sec:elem_ops_impl}
Both HAdd and PMul are fundamentally element-wise operations. 
Since the underlying representation uses multiple-limb arithmetic for each polynomial coefficient, they operate independently across both ciphertext components and limbs. On PIM, we distribute each limb of every ciphertext component (i.e., \(\llbracket m \rrbracket = c_0,c_1\)) across different PIM cores. Each PIM core hosts $N \cdot q_i$ bits (e.g., $65536 \times 8$ bytes) for one limb of one ciphertext. Given \(L\) limbs per polynomial and \(T\) total ciphertexts, we require \(2 \times L \times T\) PIM cores. The additions are performed in parallel without inter-core communication, and no alignment or synchronization is required. PMul follows the same element-wise structure as HAdd.


\subsubsection{Ciphertext-Ciphertext Multiplication (CMul)}
\label{sec:CMul_impl}
CMul involves a more complex operation: a multiplication between two ciphertexts \(\llbracket m_1 \rrbracket = c_0,c_1\) and \(\llbracket m_2 \rrbracket = c_0',c_1'\), producing three components \(d_0 = c_0 \cdot c_0'\), \(d_1 = c_0 \cdot c_1' + c_1 \cdot c_0'\), and \(d_2 = c_1 \cdot c_1'\). Each operand component is represented in multiple limbs, and we assume that all limbs for \(c_0\), \(c_1\), \(c_0'\), and \(c_1'\) are distributed across distinct PIM cores. The multiplications for \(d_0\), \(d_1\), and partial \(d_2\) terms are performed in parallel: \(c_0 \cdot c_0'\), \(c_1 \cdot c_0'\), \(c_0 \cdot c_1'\), and \(c_1 \cdot c_1'\) are each assigned to separate groups of PIM cores. Within each group, limb-level parallelism is exploited using multiple PIM cores. Once intermediate products are generated, inter-PIM core data transfer is initiated to collect terms required to construct \(d_2\). The partial results are summed locally to finalize the ciphertext multiplication. Each multiplication operation (e.g., \(c_0 \cdot c_0'\)) is a polynomial multiplication and uses NTT, element-wise multiplication (called the Hadamard product), and INTT. 

\subsubsection{Parallel NTT on PIM}
\label{sec:NTT_impl}
\begin{figure}[h]
    \centering
    \includegraphics[width=\linewidth]{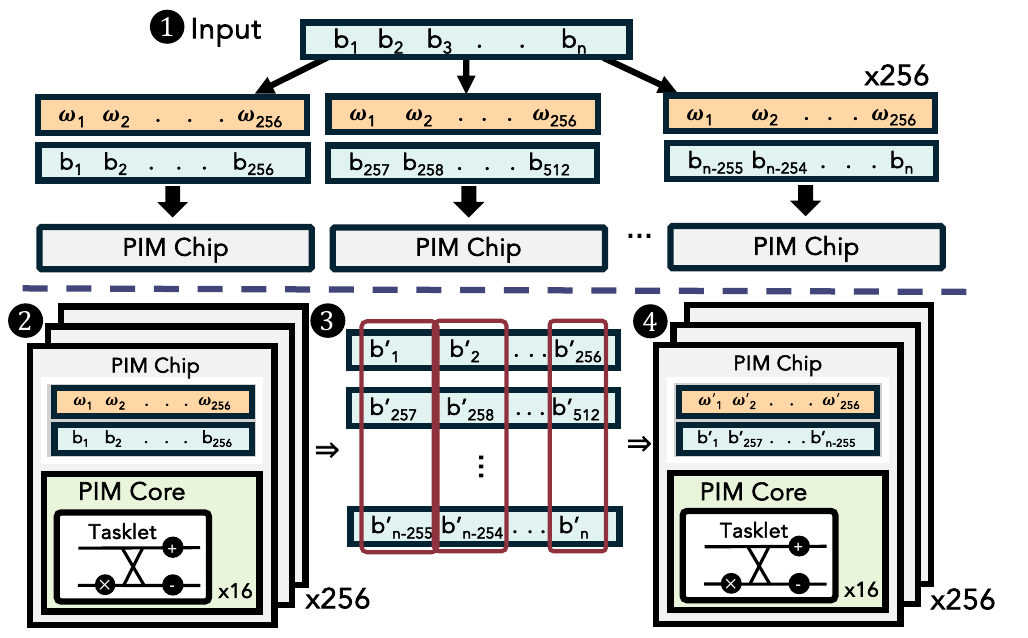}
    \caption{Dataflow and mapping of NTT on the PIM system for a single limb of a single ciphertext component.}
    \label{fig:ntt_on_PIM}
\end{figure}

Figure~\ref{fig:ntt_on_PIM} shows the mapping of NTT and INTT on a PIM system. To accelerate the Number Theoretic Transform (NTT) on PIM, prior works (e.g.,~\cite{satriawan2023conceptual,satriawan2024complete,park2023ntt,barik2026long}) exploit the parallelism in the butterfly operation by decomposing the transform into a series of modular butterfly operations, which are parallelized across PIM cores at both the coefficient and limb levels. Given a polynomial of degree \(N\), each NTT consists of \(\log_2 N\) stages, where each stage performs modular arithmetic on pairs of coefficients with twiddle factor multiplication. To map this structure efficiently to PIM, we distribute coefficients across PIM cores such that each PIM core is responsible for a fixed subset of the polynomial’s RNS limbs~\circled{1}. Within each PIM core, butterfly operations are performed independently across SIMD slots~\circled{2}. At each stage, twiddle factors are either preloaded or streamed to the PIM cores, avoiding data-movement bottlenecks. For large polynomial degrees such as \(N = 2^{16}\), we decompose the NTT into multiple blocks of coefficients and span them across additional PIM cores~\circled{3}. When coefficients span multiple PIM cores, partial results are exchanged via inter-PIM-core communication~\circled{4}. Hence, multiple PIM cores form a group for a single NTT, and each limb of the ciphertext spans multiple such PIM groups. 



\subsection{Homomorphic Subroutines on PIM}
\label{sec:he-pim-subroutines}


\subsubsection{Key Switching}
\label{sec:key_swi_impl}
Key switching is used to restore the ciphertext to its original key structure (\S\ref{sec:subroutines}). For example, after homomorphic multiplication, \emph{relinearization} is applied to reduce the ciphertext size from three components back to two by using relinearization keys to eliminate the \(d_2\) component. This requires a key-switching subroutine that transforms the \(d_2\) term into a two-component ciphertext \((d_2^{0}, d_2^{1})\) under the original secret key. Specifically, the \(d_2\) component is multiplied with the relinearization keys (a type of evaluation key), yielding \((d_2^{0}, d_2^{1}) =
(d_2 \cdot evk_0,\, d_2 \cdot evk_1)\), which themselves are encoded in the extended modulus (\(P \cdot Q\)-domain), and the resulting terms are aggregated and homomorphically added to the \(d_0\) and \(d_1\) components to produce the relinearized ciphertext.

To enable this, we perform a \textsf{ModUp} operation, which requires a basis conversion (BConv) operation that expands the modulus of the ciphertext from \(Q\) to \(P \cdot Q\), where \(P\) is a special prime basis used to facilitate key switching under the Residue Number System (RNS). This conversion is necessary because the evaluation keys are defined over the extended \(P\cdot Q\) basis, while \(d_2\) exists only in the original \(Q\) basis. 
The \textsf{ModUp} procedure consists of three steps. First, we apply the inverse Number Theoretic Transform (INTT) to bring the polynomial coefficients from the NTT domain back into the coefficient (or time) domain (\S\ref{sec:NTT_impl}). 
NTT enables efficient polynomial multiplication by transforming polynomial operands into a pointwise representation, where multiplication reduces to element-wise products. 
Second, we perform a basis extension using a precomputed BConv matrix that maps each coefficient from the \(Q\)-basis to the extended \(P \cdot Q\)-basis. This procedure increases the number of RNS limbs, allowing compatibility with the modulus size used in the relinearization. Each coefficient is lifted from \(Q\) to \(P\cdot Q\) independently, enabling fine-grained parallelism.

\begin{figure}[h]
    \centering
    \includegraphics[width=1\linewidth]{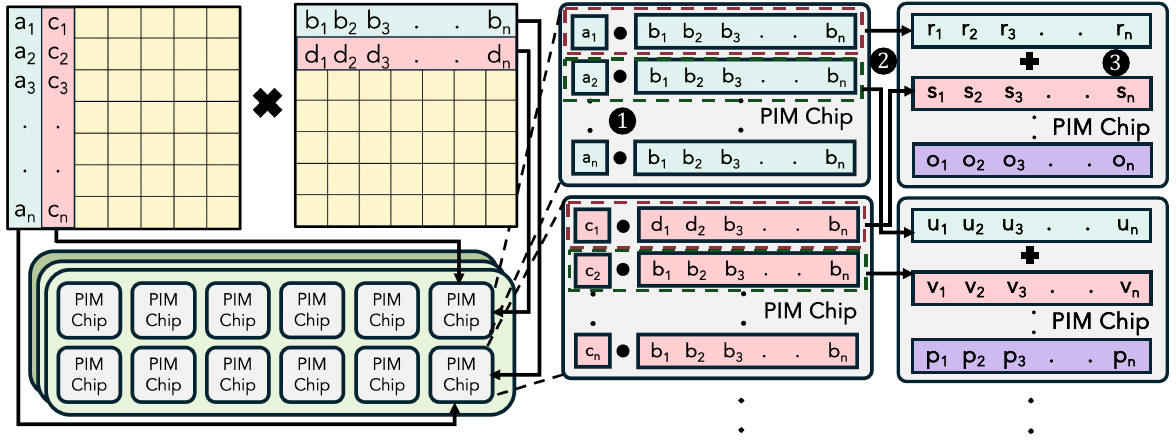}
    \caption{Data mapping of matrix multiplication on the PIM system.}
    \label{fig:matrixmul}
\end{figure}

On PIM, each limb of the \(d_2\) ciphertext component and each segment of the relinearization key are distributed across PIM cores. Figure~\ref{fig:matrixmul} shows the data mapping of matrix multiplication on the PIM system. For the BConv matrix multiplication, which maps coefficients from the \(Q\)-basis to the \(P \cdot Q\)-basis, we avoid storing the full transformation matrix on any single PIM core. Instead, each PIM core holds a single ciphertext limb and a single column of the BConv matrix to perform scalar-vector multiplication~\circled{1} against all relevant coefficients of a given limb. This approach maximizes data reuse, reduces redundant matrix storage in PIM cores, and enables multiple partial outputs to be generated in parallel across PIM cores.
After the scalar-vector multiplications are completed, the intermediate results are transferred across PIM cores~\circled{2} and aggregated~\circled{3} to produce complete RNS limbs under the \(P\cdot Q\)-basis. Each limb is present in a PIM core, and each PIM core receives multiple scale factors (as shown in Figure~\ref{fig:matrixmul}) extracted from the BConv matrix. 
This design allows us to fully utilize available parallelism while ensuring that intermediate buffers within each PIM core do not exceed the local buffer size. 
Finally, we apply the forward NTT to return the polynomial to the transform domain. 

\subsubsection{Bootstrapping}
\label{sec:boot_pim_impl}
Bootstrapping is the most complex primitive in CKKS and is required to refresh ciphertexts once the available noise budget is exhausted. The full bootstrapping procedure consists of the following stages executed in sequence: (1) modulus raising (\textsf{ModUp)}, (2) slot-to-coefficient conversion (\textsf{S2C}), (3) function evaluation (\textsf{EvalMod}), (4) coefficient-to-slot conversion (\textsf{C2S}), and (5) modulus reduction (\textsf{ModDown}).

Steps 1 and 5 are the same as those discussed above (\S\ref{sec:key_swi_impl}).
Once the ciphertext modulus is changed from \(Q\) to \(P \cdot Q\), we perform \textsf{S2C} (step 2), which extracts encoded values from the polynomial domain into the SIMD slot representation using an inverse Discrete Fourier Transform (IDFT). The IDFT matrix is precomputed and is part of this operation. This operation requires multiplying the ciphertext by the IDFT matrix (in the plaintext domain), which is the same as performing plaintext matrix-ciphertext multiplication in the $Conv$ layer. Step 4 (\textsf{C2S}) performs the inverse transformation using a DFT matrix and is structurally identical. Hence, both steps 2 and 4 use the same $\mathsf{Conv}$ kernel as \S\ref{sec:he_nn_impl}.

After \textsf{ModUp} and \textsf{S2C}, polynomial function evaluation \textsf{EvalMod} is performed over the slots. In standard CKKS bootstrapping, \textsf{EvalMod} approximates the modular reduction function using a low-degree Chebyshev polynomial to refresh the ciphertext's noise budget. When activation-bootstrapping fusion is employed~\cite{li2024evalcomp,alexandru2025general}, the activation function (e.g., square, ReLU, or GELU) can be composed with the modular reduction polynomial and evaluated jointly within this step. We represent the polynomial as a product of linear factors, \(f(x) = (x-a_1)(x-a_2)\dots(x-a_n)\), and evaluate it homomorphically using a divide-and-conquer strategy. Figure~\ref{fig:matrixmul} shows the data layout for \textsf{EvalMod} on multiple PIM cores. Each PIM core contains one limb of the ciphertext, so we divide the PIM cores into multiple groups, each containing the same ciphertext. Within a group, plaintext subtraction generates pairs of partial values, such as \(x-a_1, x-a_2\) on one PIM core and \(x-a_{i-1}, x-a_i\) on another, where $i$ is the maximum number of pairs that can be generated in parallel. These pairs are then homomorphically multiplied using CMul (\S\ref{sec:CMul_impl}) and relinearized (\S\ref{sec:key_swi_impl}). This structure allows balanced multiplication trees, minimizing multiplicative depth and enabling efficient parallel evaluation across PIM cores.

\section{Evaluation}
\label{sec:evaluation}

\subsection{Experimental Evaluation Details}
\label{sec:parameters}

We explain our evaluation in four stages. First, we decompose NN inference into three components: (i) the \(\mathsf{Conv}\) layer, implemented via \(\mathsf{PMul}\) between a Toeplitz matrix and rotated ciphertexts \(\mathsf{HRot}\);\footnote{Note that $\mathsf{S2C}$ and $\mathsf{C2S}$ are structurally analogous to the $\mathsf{Conv}$ layer because both are linear transforms implemented via plaintext matrix--ciphertext multiplication. Characterizing these kernels, therefore, captures end-to-end performance for general HE workloads.} (ii) \(\mathsf{BConv}\), the basis conversion step used in key switching when rotating ciphertexts for the \(\mathsf{Conv}\) layer or in \(\mathsf{CMul}\); and (iii) \(\mathsf{EvalMod}\), the evaluation of a Chebyshev polynomial that approximates the modular reduction function (and, when fused, the activation function) using interleaved \(\mathsf{CMul}\) and \(\mathsf{BConv}\) subroutines (\S\ref{subsec:eval_nn}). Second, we compare the UPMEM PIM implementation with a GPU baseline (\S\ref{sec:gpu}). Third, we compare two end-to-end neural networks implemented on the UPMEM PIM system with the CPU baseline (\S\ref{sec:eval_end-nn}). Fourth, we quantitatively discuss the implications for current and future PIM designs (\S\ref{sec:discussion}). 

\textbf{Parameter Selection.}
We evaluate several CKKS parameter configurations that represent different trade-offs across polynomial degree, modulus size, and multiplicative depth~\cite{ckksparams,Lattigo}. Table~\ref{tab:ckks-params} summarizes the evaluated parameter sets. These parameter sets reflect realistic deployment settings under standard security assumptions ($\geq$128-bit classical security, validated against the Homomorphic Encryption Security Standard~\cite{HomomorphicEncryptionSecurityStandard} using the parameter recommendations from the Lattigo library~\cite{Lattigo}). We use the polynomial modulus degree \(N = 2^{16}\), which is the most common degree in practical CKKS deployments for applications such as neural network inference. 
For \(N = 2^{16}\), we construct a modulus chain \(Q = \{q_0, q_1, \dots, q_k\}\) composed of prime moduli.
The modulus chains include segments allocated for homomorphic multiplication, rescaling, and transformations such as slot-to-coefficient (S2C), coefficient-to-slot (C2S), and EvalMod, which are required for CKKS bootstrapping (see \S\ref{subsec:ckks} for more details).

\begin{table}[h]
\scriptsize
\centering
\caption{CKKS parameter sets used in evaluation.}
\begin{tabular}{c | c | l}
\toprule
\textbf{$N$} & \textbf{$|Q|$} & \textbf{$Q : \{q_1, q_2, \ldots, q_n\}$} \\
\midrule
26 limbs: $2^{16}$ & 1547 & $\{60, 45\times4, 42\times3, 60\times11, 58\times4\}$ \\
27 limbs: $2^{16}$ & 1553 & $\{55, 60\times6, 30\times2, 55\times8, 53\times3\}$ \\
28 limbs: $2^{16}$ & 1767 & $\{60, 40\times3, 39\times3, 60\times9, 56\times4\}$ \\
29 limbs: $2^{16}$ & 1788 & $\{60, 45, 42\times3, 60\times11, 58\times4\}$ \\
30 limbs: $2^{16}$ & 1793 & $\{55, 60\times1, 30\times3, 55\times8, 53\times3\}$ \\
\bottomrule
\end{tabular}
\label{tab:ckks-params}
\end{table}

\textbf{Dataset Sizes.}
To evaluate the scalability and performance characteristics of homomorphic operations across varying workload intensities, we consider input datasets comprising images with spatial resolutions ranging from \(32 \times 32\) to \(512 \times 512\). 
By systematically increasing the input dimension, we capture how different HE operations scale with respect to ciphertext packing density, number of ciphertexts required, and the overall computational latency and memory footprint.

\textbf{Functional Verification.} We verify the functional correctness of all PIM-based homomorphic operations by decrypting the output ciphertexts and comparing against plaintext computation results. All operations produce outputs within the expected CKKS approximation-error bounds, confirming that our PIM implementation preserves the numerical-correctness guarantees of the CKKS scheme.

\subsection{Performance Analysis Against CPU Baselines.}
\label{subsec:eval_nn}

\subsubsection{Evaluation of Convolution Layers ($\mathsf{Conv}$)}
\label{sec:conv}

\begin{figure}[h]
    \centering
    \includegraphics[width=\linewidth]{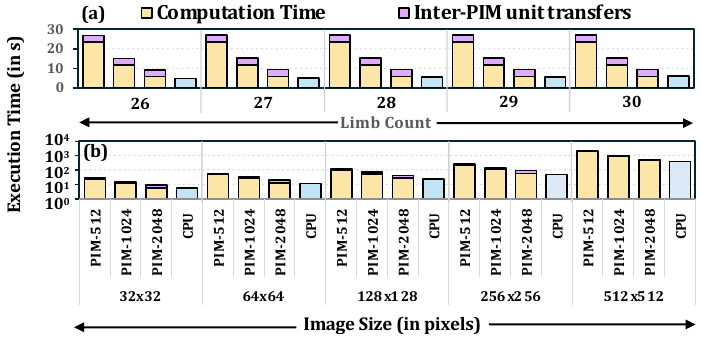}
    \caption{Total execution time (in s) for the $\mathsf{Conv}$ layer on the UPMEM PIM system for (a) varying limb count and (b) varying image sizes compared with the CPU system.}
    \label{fig:conv}
\end{figure}

\textbf{Figure~\ref{fig:conv}(a)} shows the execution time of the $\mathsf{Conv}$ layer on our PIM system as a function of two parameters: the number of DPUs and the HE parameter choices (expressed as ciphertext limb count).
The input image size is fixed to $32\times32$. 
We make four observations (\#1-\#4). 

\observation{For a fixed number of PIM cores, the execution time of the $\mathsf{Conv}$ layer remains similar across different ciphertext limb counts.}

For example, the execution times of the $\mathsf{Conv}$ operation with limb counts of 26 and 30 are similar. Although increasing the limb count increases the number of plaintext–ciphertext operations from 1600 to 1920, this difference does not affect end-to-end latency. For a fixed PIM configuration, the number of execution rounds remains constant (four rounds for 512 PIM cores and a single round for 2048 PIM cores) regardless of the chosen limb count.

\observation{Increasing the number of PIM cores significantly reduces execution time across all ciphertext limb counts.}

For example, the execution time of the $\mathsf{Conv}$ operation on 2048 PIM cores is $2.8\times$ lower than on 512 cores. 
As HE exposes substantial limb-level parallelism, 
it maps efficiently to the increased parallel compute capacity of PIM configurations.

\observation{The execution time of the $\mathsf{Conv}$ operation on the PIM system is dominated by the cost of the aggregation step.}

For example, our breakdown of total execution time shows that $\sim$78\% of the time is spent on aggregating rotated ciphertexts. In comparison, plaintext-ciphertext multiplication accounts for only 9\% of the total time, and data transfers between PIM cores and the host CPU for data layout management account for 12\%. This is because all computations are performed in modular arithmetic, where the modulo operation (performed using Barrett reduction~\cite{barrett1987implementing}) dominates the arithmetic cost and accounts for 96–99\% of the total computation time. As the aggregation phase repeatedly invokes Barrett reduction, it becomes the primary contributor to overall execution time.

\observation{The data-transfer time between the PIM cores and the host CPU increases with both the number of PIM cores and the size of the ciphertext limb count.}

For example, the fraction of total execution time spent on data transfer increases from 12\% to 35\% as the number of PIM cores increases from 512 to 2048 at 26 limbs. Similarly, for a fixed configuration of 2048 PIM cores, increasing the limb count from 26 to 30 increases the transfer overhead from 35\% to 39\%.

\textbf{Figure~\ref{fig:conv}(b)} shows the execution time of the $\mathsf{Conv}$ operation on our PIM system as a function of two factors: the number of DPUs and the input image size. The limb count is fixed to 30. We make three observations (\#5-\#7).

\observation{For a fixed number of PIM cores, doubling the image size doubles the execution time of the $\mathsf{Conv}$ operation. However, once the image size exceeds the ciphertext size, performance degrades significantly.}

For example, on a 512-core PIM system, the execution time for a $32\times32$ image is $27.33$\,s, whereas increasing the image size to $64\times64$ increases the execution time to $58.7$\,s. This scaling occurs because increasing the image size increases the number of required ciphertext rotations in proportion to $\sqrt{I}$ (\S\ref{sec:he_nn_impl}), resulting in more rotated ciphertexts that must be evaluated and, hence, more PIM-core work.

However, when the image size grows beyond the ciphertext capacity, for instance, to $512\times512$ (requiring $262{,}144$ slots), performance degrades sharply. The total execution time for a $512\times512$ image is $1967.8$\,s, which is $71.9\times$ higher than for a $32\times32$ image. This slowdown stems from two factors: (1) the larger image requires multiple ciphertexts to represent the input, increasing the total number of ciphertexts that must be processed, and (2) the number of ciphertext rotations grows with $\sqrt{I}$, further increasing the overall computational time.

\observation{Increasing the number of PIM cores reduces the execution time of the $\mathsf{Conv}$ kernel.}

For example, the execution time decreases from $27.3$\,s to $15.6$\,s and $9.7$\,s as the number of PIM cores increases from 512 to 1024 and 2048, respectively. This is because the homomorphic operations in the $\mathsf{Conv}$ layer exhibit substantial parallelism that additional PIM cores can directly exploit. As a result, doubling or quadrupling the number of PIM cores yields a corresponding $2$-$4\times$ reduction in computation time.

\observation{For smaller data sizes, the execution time on large PIM-core configurations is limited by both data-transfer overheads and computation. In contrast, for larger data sizes, execution time is dominated solely by the computation phase.}

For instance, (1) on smaller PIM configurations (512 PIM cores), data-transfer overhead accounts for 12-14\% of the total execution time for image sizes ranging from $32\times32$ to $256\times256$, but this fraction drops to only 3\% for a $512\times512$ image. 
(2) For larger PIM configurations (2048 PIM cores), transfer time constitutes 34-40\% of the total execution time for $32\times32$ to $256\times256$ images, while for a $512\times512$ image it decreases to 11\%. 
This trend arises because the $\mathsf{Conv}$ operations on larger images are dominated by computation, particularly the Barrett-reduction~\cite{barrett1987implementing} operation. Hence, the aggregation phase (which is computation-heavy) becomes the primary contributor to end-to-end latency for large image sizes.

\subsubsection{Evaluation of Basis Conversion ($\mathsf{BConv}$)}
\label{sec:basis_conv}

\begin{figure}[h]
    \centering
    \includegraphics[width=\linewidth]{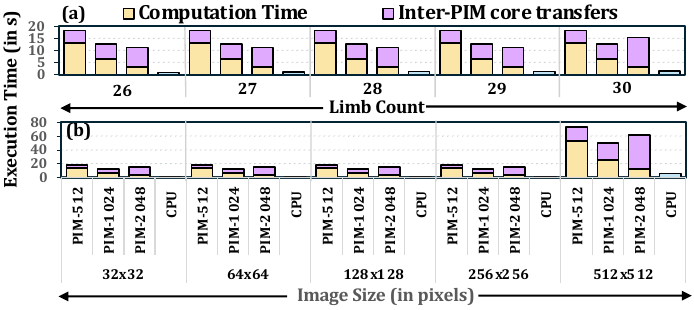}
    \caption{Total execution time (in s) for the $\mathsf{BConv}$ subroutine on the UPMEM PIM system for (a) varying limb count and (b) varying image sizes compared with the CPU system.}
    \label{fig:bconv}
\end{figure}

\textbf{Figure~\ref{fig:bconv}(a)} shows the execution time of the $\mathsf{BConv}$ operation on our PIM system as a function of two parameters: the number of DPUs and the HE parameter setting (expressed as ciphertext limb count). The input image size is fixed to $32\times32$. 
We make the following observation (\#8).

\observation{As the limb count increases, the execution time of the $\mathsf{BConv}$ subroutine increases due to the increased latency of $\mathsf{NTT}$ and $\mathsf{INTT}$ operations.}

The execution time of matrix multiplication in the $\mathsf{BConv}$ subroutine is constant because the twiddle-factor matrix size depends only on the polynomial degree ($N$) and not on the limb count. Hence, the number of PIM execution rounds remains the same. However, the total time spent in the $\mathsf{NTT}$ operations (required in the $\mathsf{BConv}$ kernel) increases from $64.1$\,s to $76.8$\,s as the limb count grows. This increase arises because each limb utilizes 256 PIM cores to maximize parallelism for the butterfly computation. Hence, a higher limb count directly increases both the number of execution rounds and the associated data-transfer overhead.

\textbf{Figure~\ref{fig:bconv}(b)} shows the execution time of the $\mathsf{BConv}$ operation on our PIM system as a function of two factors: the number of DPUs and the input image size. The limb count is fixed to 30. We make three observations (\#9-\#11).

\observation{For a fixed number of PIM cores, increasing the image size does not change the total execution time of the $\mathsf{BConv}$ subroutine. However, once the image size exceeds the ciphertext size, the execution time significantly increases.}

For example, the latency of each kernel (matrix multiplication, $\mathsf{NTT}$, and $\mathsf{INTT}$) is independent of the image size. This is because the $\mathsf{BConv}$ routine depends only on the number of ciphertexts for a fixed limb count. As a result, the total latency remains nearly constant across image sizes (e.g., $18.1$\,s for 512 PIM cores). However, when the image size exceeds the ciphertext size (e.g., $512\times512$), a single ciphertext cannot hold the entire input, the number of ciphertexts increases, and the total execution time rises from $18.1$\,s to $73.0$\,s.

\observation{Data transfer between PIM cores and the host CPU is a major bottleneck in the $\mathsf{BConv}$ routine.}

For example, the total execution time decreases significantly as the number of PIM cores increases from 512 to 1024, but the improvement from 1024 to 2048 cores is comparatively modest. This is because data transfers between the PIM cores and the host CPU dominate the matrix-multiplication phase. For a $512\times512$ image, the computation and transfer times are $52.6$\,s and $20.3$\,s, respectively, on a 512-core configuration, whereas they are $13.1$\,s and $48.5$\,s on a 2048-core configuration. Thus, the data movement required across multiple PIM rounds becomes the primary bottleneck, limiting scalability beyond a certain core count.

\observation{Increasing the number of PIM cores does not allow the PIM system to outperform the CPU system for the $\mathsf{BConv}$ subroutine because of the data-transfer bottleneck.}
This is because, for the $\mathsf{BConv}$ subroutine, increasing the number of PIM cores also increases data movement. This overhead arises from data layout management and the need to perform data transposition during matrix-matrix multiplication (\S\ref{sec:key_swi_impl}).

\subsubsection{Evaluation of Polynomial Approximation ($\mathsf{EvalMod}$)}
\label{sec:evalmod}

\begin{figure}[h]
    \centering
    \includegraphics[width=\linewidth]{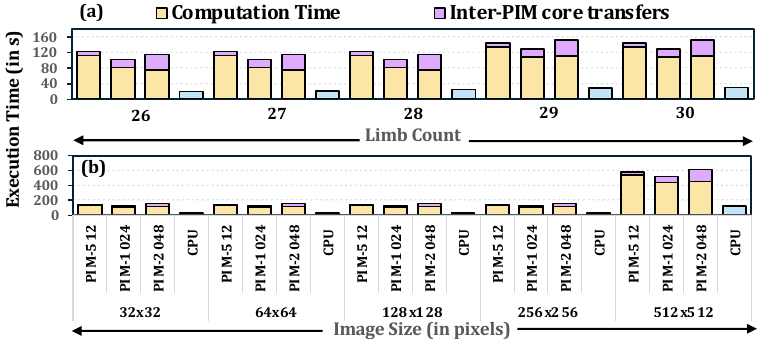}
    \caption{Total execution time (in s) for the $\mathsf{EvalMod}$ operation on the UPMEM PIM system for (a) varying limb count and (b) varying image sizes compared with the CPU system.}
    \label{fig:polyeval}
\end{figure}

\textbf{Figure~\ref{fig:polyeval}(a)} shows the execution time of the $\mathsf{EvalMod}$ operation on our PIM system as a function of two factors: the number of DPUs and the HE parameter choices. The input image size is fixed to 32$\times$32. 
We make two observations (\#12-\#13). 

\observation{For a fixed number of PIM cores, increasing the number of limbs increases the execution time of $\mathsf{EvalMod}$.}

For example, on a 512-core PIM system, the total execution time for the $\mathsf{EvalMod}$ operation increases from $100.7$\,s at a limb count of 25 to $144.8$\,s at a limb count of 30. A similar trend holds for the 2048-core PIM configuration, where execution time increases from $115.9$\,s to $152.8$\,s as the limb count increases from 25 to 30. This increase occurs because a higher limb count requires more PIM execution rounds, thereby increasing the overall latency.

\observation{The execution time of the PIM system varies substantially with the number of PIM cores used.}

For example, for a limb count of 25, the total execution time of the $\mathsf{EvalMod}$ routine is $100.7$\,s on a 512-core PIM system and $115.9$\,s on a 2048-core system. For a limb count of 30, the execution time is $144.8$\,s on 512 PIM cores, $129.7$\,s on 1024 cores, and increases to $152.8$\,s on 2048 PIM cores. This is because increasing the number of PIM cores does not substantially reduce the computation time of $\mathsf{EvalMod}$, while the associated data-transfer overhead grows with core count. The combined effect of limited computational speedup and increasing transfer overhead determines the performance tradeoff, indicating that $\mathsf{EvalMod}$ is largely compute-bound and does not benefit significantly from additional parallel units; synchronization and communication costs instead degrade performance at higher PIM-core counts.

\textbf{Figure~\ref{fig:polyeval}(b)} shows the execution time of the $\mathsf{EvalMod}$ operation on our PIM system as a function of two factors: the number of DPUs and the input image size. The limb count is fixed to 30. Similar to the $\mathsf{BConv}$ routine, increasing the image size does not affect the execution time for a fixed number of PIM cores until the image size exceeds the ciphertext capacity.
We make two observations (\#14-\#15).

\observation{Computation time is a significant bottleneck in the $\mathsf{EvalMod}$ operation.}

For example, on a 512-core PIM system, computation accounts for approximately 93\% of the total execution time across all image sizes when a single ciphertext is sufficient, and this fraction increases to 97\% when multiple ciphertexts are required. On a 2048-core system, computation constitutes about 76\% of the total execution time with a single ciphertext and up to 85\% when multiple ciphertexts are involved. This is because the $\mathsf{EvalMod}$ routine is dominated by extensive element-wise operations, which impose substantial computational cost regardless of the image size.

\observation{The CPU outperforms PIM for the $\mathsf{EvalMod}$ operation.}

This is because the $\mathsf{EvalMod}$ operation uses the same ciphertext to perform the polynomial approximation. As a result, the CPU exploits very high data reuse, whereas our PIM system lacks computational power due to the absence of a 64-bit modular hardware multiplier unit.  

\subsection{Performance Analysis Against GPU Baselines.}
\label{sec:gpu}

\begin{figure}[h]
    \centering
    \includegraphics[width=\linewidth]{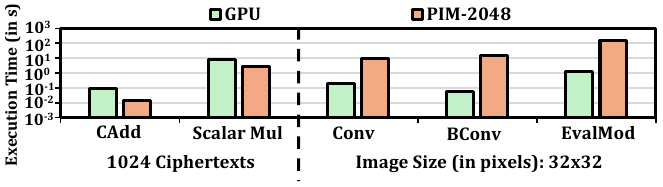}
    \caption{Total execution time (in s) for all operations/subroutines on the UPMEM PIM system compared with a GPU system.}
    \label{fig:gpu}
\end{figure}

\textbf{Figure~\ref{fig:gpu}} shows the execution time (in s) of element-wise operations ($\mathsf{CAdd}$, $\mathsf{Scalar Mul}$) on two vectors of 1024 ciphertexts, and three homomorphic operations/subroutines ($\mathsf{Conv}$, $\mathsf{BConv}$, $\mathsf{EvalMod}$) for a 32$\times$32 image. 
We make three observations (\#16-\#18). 

\observation{The PIM system outperforms the GPU system for $\mathsf{element\text{-}wise}$ operations (CAdd, Scalar Mul).}
This is because the $\mathsf{element\text{-}wise}$ operations (CAdd, ScalarMul) can be highly parallelized and require no data transfer between the PIM cores. In contrast, GPUs move data from the host's main memory to GPU memory and then to the compute units. Gupta et al.~\cite{gupta2023evaluating} report a similar observation, and our results are consistent with theirs.

\observation{The GPU significantly outperforms the PIM system for $\mathsf{Conv}$, $\mathsf{BConv}$, and $\mathsf{EvalMod}$.}
(1) For the $\mathsf{Conv}$ layer on a $32\times32$ input with a limb count of 30, the GPU achieves an execution time of $0.2$\,s, whereas the PIM system (2048 cores) requires $9.3$\,s. 
(2) For the $\mathsf{BConv}$ subroutine, the GPU executes the computation in $0.06$\,s compared to $15.4$\,s on the PIM system.
(3) For the $\mathsf{EvalMod}$ operation, the GPU execution time is $1.2$\,s, while the PIM system requires $152.8$\,s.  

This is due to two key factors. First, GPUs provide native 64-bit integer multipliers. In contrast, UPMEM PIM cores expose only 8-bit multipliers and emulate 64-bit arithmetic via a sequence of shift-and-multiply micro-operations, resulting in significantly higher latency per multiplication. 
Second, modern GPUs incorporate specialized multiply-and-accumulate (MAC) units, which directly accelerate homomorphic operations such as $\mathsf{HMul}$, $\mathsf{HRot}$, and the matrix-multiplication steps in $\mathsf{BConv}$.

\observation{The GPU system cannot process large image sizes because of out-of-memory constraints when using a single-GPU configuration.}

Although GPUs offer high throughput, their HBM capacity (typically 40–80 GB) can limit large FHE workloads. On a single A100, inputs beyond 256$\times$256 exceed memory because $\mathsf{Conv}$ requires a Toeplitz matrix and many rotated ciphertexts whose footprint scales rapidly with image size; deeper networks (e.g., AlexNet/ResNet) further amplify this via intermediate ciphertext accumulation. We note that this out-of-memory limitation may also be partially attributable to the memory management of the specific library used (NEXUS~\cite{cryptoeprint:2024/136}); other implementations with more memory-efficient strategies may support somewhat larger inputs on the same hardware. Consequently, prior GPU-based FHE studies typically evaluate only small inputs (e.g., 32$\times$32)~\cite{cuFHE}. While multi-GPU execution could mitigate capacity limits, it requires partitioning ciphertexts and computations across devices and introduces inter-GPU communication via PCIe~\cite{pcie-2017} or NVLINK~\cite{li2019evaluating}. Since current FHE libraries do not support multi-GPU execution, we leave this to future work.

\subsection{Evaluation of End-to-End Neural Networks.}
\label{sec:eval_end-nn}

\begin{figure}[h]
    \centering
    \includegraphics[width=\linewidth]{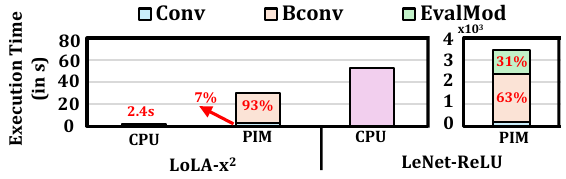}
    \caption{Execution time (in s) for end-to-end neural networks.}
    \label{fig:end-to-end-nn}
\end{figure}

We compare the performance of two end-to-end neural networks (LoLA-$x^{2}$~\cite{LoLa19}, LeNet-ReLU~\cite{lecun1998gradient}) for a single MNIST image of size 32$\times$32$\times$3. For LoLA, we use 5 limbs with $N=2^{16}$, and for LeNet, we use 30 limbs with $N=2^{16}$. \textbf{Figure~\ref{fig:end-to-end-nn}} shows the execution time (in s) for the two networks. We observe that the CPU system is 16.6$\times$/64.0$\times$ faster than the HE-based NN implementation (LoLA-$x^{2}$/LeNet-ReLU) on existing UPMEM-PIM systems, due to slower individual kernels/subroutines. LoLA-$x^2$ is a simple network that does not require bootstrapping and spends 7\% and 93\% of the total execution time in $\mathsf{Conv}$ and $\mathsf{BConv}$, respectively. LeNet-ReLU involves a complex activation function, which requires bootstrapping and spends 6\%, 63\%, and 31\% of the total time in $\mathsf{Conv}$, $\mathsf{BConv}$, and $\mathsf{EvalMod}$.\footnote{Note that we consider the latency of $\mathsf{S2C}$ and $\mathsf{C2S}$ in $\mathsf{Conv}$.}

\subsection{Discussion on Future PIM Systems for HE}
\label{sec:discussion}

We identify key limitations of current PIM platforms and motivate an HE-optimized future PIM system, which we model and compare against CPU/GPU baselines. Overall, existing PIM systems underperform CPUs and GPUs due to two limitations: (i) the absence of wide (e.g., 64-bit) hardware modular arithmetic such as Barrett multiplication~\cite{barrett1987implementing,chmielewski2025upmem}, and (ii) the lack of direct communication links between PIM cores~\cite{gomez2021benchmarking,gomez2022benchmarking,hyun2024pathfinding,jonatan2024scalability}. These constraints are also common in other PIM designs (e.g., Samsung Aquabolt~\cite{kim2021aquabolt}, SK Hynix AiM~\cite{kwon2022system}), which similarly lack modular units and hardware interconnect support.


\textbf{1) Improving Arithmetic Throughput in PIM Cores.} The computational throughput of current PIM cores is fundamentally limited by the absence of a native 64-bit integer modular multiplier. We observe that each PIM core spends 99\% of its time performing modulo operations (i.e., Barrett reduction) for modular addition (used in $\mathsf{HAdd}$) and 96\% for modular multiplication (used in $\mathsf{PMul}$, $\mathsf{CMul}$, and other subroutines). Adopting a hardware-based Barrett multiplier can eliminate the need for costly software emulation of modular operations. 

\textbf{2) Support for Interconnects.} Our analysis (\S\ref{sec:basis_conv}) shows that the $\mathsf{BConv}$ kernel is bottlenecked by data movement between PIM cores. This is because existing PIM systems have very limited memory capacity per PIM core and therefore require frequent communication between PIM cores~\cite{gomez2021benchmarking,gomez2022benchmarking,jonatan2024scalability}. In contrast, state-of-the-art GPU systems treat GPU memory as a unified address space (e.g., a single 40 GB or 80 GB memory in A100 GPUs), where all compute units can access the full memory capacity. This limitation worsens when each PIM core is augmented with a modular multiplier, which improves arithmetic throughput but exacerbates the data-movement bottleneck.

\textbf{Methodology and Baselines.} We evaluate the performance potential of an HE-optimized PIM architecture using the UPMEM PIM simulator, modeling three designs:
\squishlist
\item \textbf{PIM-Barrett}: We augment each PIM core with a multi-stage 64-bit Barrett multiplier, which is pipelined to produce output every cycle. 
This configuration quantifies the benefit of \emph{only} accelerating modular arithmetic. 

\item \textbf{PIM-Barrett-Ideal-Net}: Building on PIM-Barrett, we assume an ideal, zero-latency, all-to-all interconnect among PIM cores. Under this model, any core can transfer data to another bank without communication delay, effectively treating the PIM system as a single unified memory. This configuration provides a best-case bound on performance when inter-core data movement is not a limiting factor.

\item \textbf{PIM-Barrett-Net}: We use a realistic interconnect model proposed in prior work, PIMnet~\cite{son2025pimnet}. The simulator accounts for finite bandwidth and data-transfer latency, consistent with the PIMnet design. This configuration serves as a state-of-the-art baseline for inter-DPU communication. 
\squishend

We evaluate three configurations in the UPMEM PIM simulator with 2048 cores by extending in two dimensions: (i) we add a multi-stage 64-bit Barrett modular multiplier with one modular multiplication per cycle, replacing software-emulated modulo operations, and (ii) we model inter-DPU communication as a byte-parameterized cost with zero latency (Ideal) or fixed latency/bandwidth (PIMnet). The baseline UPMEM platform has 64 DPUs per rank and a bandwidth limit of 4.74 GB/s (CPU→DPU) and 6.68 GB/s (DPU→CPU)~\cite{gomez2021benchmarking}. PIMnet enables direct bank-to-bank transfers with 179.2 GB/s aggregated per-rank send+receive bandwidth~\cite{son2025pimnet}. We execute the same HE kernels/subroutines across all configurations, record total simulated cycles, convert to latency using the modeled frequency, and compare against CPU, GPU, and the unmodified UPMEM baseline.

\begin{figure}[h]
    \centering
    \includegraphics[width=\linewidth]{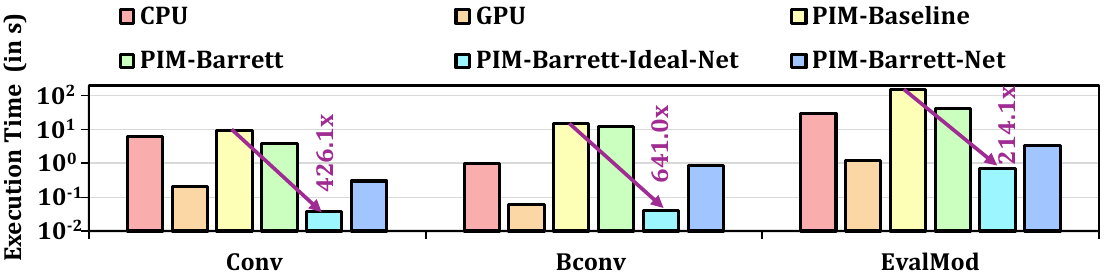}
    \caption{Total execution time on three PIM designs compared with the existing PIM system, CPU, and GPU systems.}
    \label{fig:discussion}
\end{figure}

\textbf{Figure~\ref{fig:discussion}} shows the execution time (in s) of three homomorphic operations/subroutines ($\mathsf{Conv}$, $\mathsf{BConv}$, $\mathsf{EvalMod}$) for a 32$\times$32 image size. 
We make four observations.

\squishlist 
\item First, PIM-Barrett improves modular addition/multiplication by 158.5$\times$ on average, and yields up to 2.3$\times$ speedup for $\mathsf{Conv}$, 1.26$\times$ for $\mathsf{BConv}$, and 3.6$\times$ for $\mathsf{EvalMod}$ on 32$\times$32 inputs compared to PIM-Baseline. We observe that PIM-Barrett outperforms PIM-Baseline \emph{only} for the $\mathsf{Conv}$ layer because it is bottlenecked by computations. Similarly, for $\mathsf{EvalMod}$, PIM-Barrett achieves performance comparable to the CPU; however, increasing computational capability exacerbates data-movement overheads, preventing further gains over CPU execution. Moreover, PIM-Barrett underperforms the CPU for $\mathsf{BConv}$ and the GPU across all kernels, as inter-core data movement remains the dominant bottleneck.

\item Second, PIM-Barrett-Ideal-Net provides 426.1$\times$/641.0$\times$/ 214.1$\times$ speedups over PIM-Baseline for $\mathsf{Conv}$/$\mathsf{BConv}$/ $\mathsf{EvalMod}$, respectively. It outperforms the CPU by 163.8$\times$/ 23.5$\times$/42.3$\times$ and the GPU by 5.3$\times$/1.4$\times$/1.06$\times$ across $\mathsf{Conv}$/$\mathsf{BConv}$/$\mathsf{EvalMod}$.

\item Third, for $\mathsf{EvalMod}$, PIM-Barrett-Ideal-Net is only 1.06$\times$ faster than the GPU because $\mathsf{EvalMod}$ repeatedly evaluates a polynomial on the same ciphertext via a divide-and-conquer scheme (\S\ref{sec:boot_pim_impl}), yielding high data reuse but limited inter-ciphertext parallelism. Further performance improvement, therefore, requires scaling out with more PIM cores/chips to better exploit parallelism within $\mathsf{EvalMod}$.

\item Fourth, PIM-Barrett-Net executes $\mathsf{Conv}$/$\mathsf{BConv}$/$\mathsf{EvalMod}$ in 0.29/0.85/3.45\,s, corresponding to 20.4$\times$/1.14$\times$/8.75$\times$ speedups over the CPU. $\mathsf{BConv}$ improves only marginally because it incurs frequent transfers across INTT $\rightarrow$ matmul $\rightarrow$ NTT; higher bandwidth helps, but communication frequency dominates, whereas CPUs/GPUs benefit from a unified memory space. Overall, PIM-Barrett-Net still lags the GPU due to (i) frequent inter-bank data movement driven by small bank-local memory and (ii) lower operating frequency than the GPU.    

\squishend

\textbf{Hardware Overhead of PIM-Barrett-Net.} PIM-Barrett-Net adds two hardware units to the existing PIM-Baseline: (i) a PIMnet-based interconnect~\cite{son2025pimnet} and (ii) a 64-bit modular Barrett multiplier. Prior work (PIMNet~\cite{son2025pimnet}) reports area and power overheads from synthesizing the Verilog implementation with OpenRoad 45nm technology (NANDgate45)~\cite{ajayi2019openroad} and restricts the design to three metal layers (similar to DRAM technology). The overhead of the interconnect unit is $0.013 \text{mm}^{2}$ in area and $17$\,mW in power consumption per rank. We use the same methodology to calculate the area and power overhead of a 64-bit Barrett multiplier. We design a 26-stage 64-bit Barrett multiplier~\cite{barrett1987implementing} that produces output every cycle, occupying $0.083\text{mm}^2$ and consuming $17.3$\,mW per bank. This overhead is negligible compared to the DRAM die area~\cite{shim201816gb}.

\section{Discussion}
\label{sec:final_discussion}

\textbf{Scalability.} PIM systems scale effectively for computation-dominated kernels because of their modular design. For example, 8 PIM modules provide 2560 PIM cores, and scaling to 16 or more modules provides more computational capacity. This increased parallelism can directly accelerate the $\mathsf{Conv}$ layer, as it is dominated by computations (\S\ref{subsec:eval_nn}), and more PIM modules can exploit the inherent parallelism of HE (\S\ref{sec:parallelism}). However, for communication-heavy kernels such as $\mathsf{BConv}$ and $\mathsf{EvalMod}$, scaling is limited by inter-core data movement and synchronization overheads (\S\ref{sec:basis_conv}, \S\ref{sec:evalmod}), and adding more PIM cores yields diminishing returns without improved interconnects. Scaling a PIM system is efficient in terms of both cost and power consumption compared to GPU systems~\cite{mutlu2022modern,gomez2021benchmarking,falevoz2024energy}.

\textbf{Optimal Interconnect Design.} PIM performance can be substantially improved through a better interconnect co-design. While PIMnet~\cite{son2025pimnet} adopts a hierarchical ring–crossbar–bus topology, communication in HE-based kernels is dominated by all-to-all communication and the structured patterns of NTT/INTT. Interconnects that explicitly target these patterns, e.g., by incorporating hardware-based butterfly-style permutations~\cite{park2023ntt} or leveraging rearrangeable networks such as the Beneš network~\cite{manuel2008efficient,samardzic2022craterlake}, can narrow the gap between practical designs and the PIM-Barrett-Ideal configuration.

\textbf{Optimal HE parameters.} PIM systems can efficiently scale to large limb counts (e.g., 30 in our evaluation) by leveraging limb-level parallelism across many cores. Increasing the number of limbs reduces the frequency of costly bootstrapping, thereby improving HE performance. In contrast, higher limb counts increase memory footprint and data-movement overheads on CPU/GPU platforms (\S\ref{sec:evaluation}). 
Hence, future PIM architectures can be tuned to use large limb counts, with different memory-to-compute ratios that help manage noise growth and minimize bootstrapping overhead.

\section{Related Work}
\label{sec:related}


To our knowledge, this paper is the first to implement and rigorously
evaluate all homomorphic operations and subroutines on a real-world PIM system used in modern applications. We describe other related works.

\noindent \textbf{Software-based HE Libraries:} State-of-the-art HE libraries (e.g.,~\citecat{libraries}) implement the BFV, BGV, and CKKS schemes with extensive algorithmic optimization, including optimized RNS pipelines, fused modular arithmetic, and highly tuned NTT kernels that maximize memory locality and minimize noise growth. We leave these intensive algorithm optimizations for future work.

\noindent \textbf{Hardware Acceleration.} Several accelerators and libraries for HE on compute-centric platforms (e.g., GPUs and FPGAs) have been developed. For example, Intel HEXL~\cite{hexl} provides highly optimized polynomial-arithmetic kernels for AVX-equipped CPUs~\cite{avx2011intel}. GPU-based systems (e.g., CuHE~\cite{dai2016cuhe}, CuFHE~\cite{cuFHE}, ARK~\cite{kim2022ark}) accelerate core HE kernels by exploiting high-throughput SIMD parallelism.
FPGA-based implementations (e.g.,~\citecat{fpga}) exploit reconfigurable datapaths for efficient polynomial arithmetic, but are limited by programmable-logic resources and memory-bandwidth constraints.
ASIC-based accelerators (e.g., F1~\cite{feldmann2021f1}, Craterlake~\cite{samardzic2022craterlake}, BTS~\cite{kim2022bts}, SHARP~\cite{kim2023sharp}, Cheetah~\cite{reagen2021cheetah}) achieve substantially higher throughput by leveraging large on-chip SRAM and custom parallel units. However, GPU, FPGA, and ASIC designs all face data movement bottlenecks as ciphertext sizes grow.

\noindent \textbf{Processing-in-memory (PIM).} Prior PIM-based solutions for HE primarily explore simple arithmetic kernels that can benefit from execution on parallel PIM chips. While basic operations such as ciphertext addition and coefficient-wise multiplication map well to PIM, more complex kernels (e.g., NTT and key switching) require non-local communication that exceeds the limited scratchpad and inter-bank connectivity of existing PIM systems. FheMem~\cite{zhou2025fhemem} attempts to address this challenge by adding horizontal links and inter-bank communication, but doing so requires substantial modifications to commodity HBM chips. 
Prior works (e.g.,~\cite{bpntt,dramaton,park2023ntt,barik2026long,mwaisela2024evaluating}) are limited to NTT and do not evaluate large-scale workloads. 

Existing HE accelerators do not resolve the fundamental trade-off between arithmetic throughput and data movement. GPU- and ASIC-based designs provide high computational capability but are ultimately limited by the ciphertext footprint and the associated memory-transfer overheads. In contrast, prior PIM-based approaches reduce data movement but have not demonstrated how to efficiently support the full execution pipeline of HE schemes. 
Our work bridges this gap by identifying which CKKS kernels most benefit from in-memory execution, determining where hybrid execution (PIM + host/GPU) is necessary, and quantifying the resulting performance and communication trade-offs across the end-to-end HE workflow.

\section{Conclusion}
\label{sec:conclusion}

This work presents the first comprehensive characterization of core HE operations on a \emph{real-world} general-purpose PIM system. We implement and evaluate key HE operations and subroutines on PIM and analyze their performance and scalability. 
Our results show that PIM is well-suited for memory-bound HE kernels (e.g., ciphertext addition and convolution) by exploiting large-scale parallelism. However, compute-intensive and communication-heavy kernels~(e.g., $\mathsf{BConv}$, $\mathsf{EvalMod}$) are limited by the lack of wide-precision arithmetic units and inter-DPU communication support. We hope our study provides a foundation for designing future PIM architectures that are better optimized for homomorphic workloads, and that our design recommendations can guide both hardware architects and HE software developers toward more efficient algorithm-hardware co-design.

\bibliographystyle{IEEEtran} 

\bibliography{ref-filter-dedup}

\end{document}